\documentclass{aa}  

\usepackage{graphicx}
\usepackage{xcolor}
\usepackage{caption}
\usepackage{txfonts}
\usepackage{longtable}
\usepackage{float}
\usepackage{placeins}
\usepackage{lscape}
\begin{document}

   \title{High energy time lags of Gamma Ray Bursts}

   \author{C. Maraventano
          \inst{1}
          \and
          G. Ghirlanda\inst{2}
          \and
          L. Nava\inst{2,3}
          \and 
          T. Di Salvo\inst{1}
          \and
          W. Leone \inst{4}
          \and
          R. Iaria\inst{1}
          \and
          L. Burderi \inst{5,6}
          \and 
          A. Tsvetkova \inst{6,7,8}
          }

   \institute{Dipartimento di Fisica e Chimica - Emilio Segrè, 
              Università di Palermo, via Archirafi 36 - 90123 Palermo, Italy
         \and
             INAF - Osservatorio Astronomica di Brera, Via E. Bianchi 46, I-23807, Merate (LC), Italy
          \and
          INFN, Sezione di Trieste, I-34127 Trieste, Italy
          \and
          Dipartimento di Fisica, Università di Trento, Sommarive 14, 38122 Povo (TN), Italy
          \and
          Istituto di Astrofisica Spaziale e Fisica Cosmica, Istituto nazionale di
          astrofisica, via Ugo La Malfa 153, Palermo, I-90146, Italia.
          \and 
          Department of Physics, University of Cagliari, SP Monserrato-Sestu, km 0.7, 09042 Monserrato, Italy
          \and
          INAF – Osservatorio di Astrofisica e Scienza dello Spazio di Bologna, Via Piero Gobetti 101, I-40129 Bologna, Italy
          \and
          Ioffe Institute, Politekhnicheskaya 26, 194021 St. Petersburg, Russia\\
             }

   \date{Date}

% \abstract{}{}{}{}{} 
% 5 {} token are mandatory
 
\abstract
  % Context heading (optional)
   {Positive lags between the arrival time of different photon energies are commonly observed in the prompt phase of Gamma-Ray Bursts (GRBs), where soft photons lag behind harder ones. However, a fraction of GRBs display the opposite behavior. In particular, \emph{Fermi} Large Area Telescope (LAT) observations revealed that high-energy photons are often characterized by a delayed onset.}
  % Aims heading (mandatory)
   {We explore the potential of spectral lags as a diagnostic tool to identify distinct emission components or processes. By analyzing data from the \emph{Fermi} Gamma-ray Burst Monitor (GBM) and the LAT Low Energy (LLE) technique, we explore the connection between lag behavior and high-energy spectral properties.}
  % Methods heading (mandatory)
   {We analyze a sample of 70 GRBs from the LLE Catalog. Spectral lags are computed using the Discrete Correlation Function method, considering light curves extracted in four different energy bands, from 10 keV to 100 MeV. Additionally, we compare LLE time lags with properties of the prompt emission and with the spectral behavior at high energies.}
  % Results heading (mandatory)
   {Time lags computed across different energy bands distributed between 10 keV and 1 MeV are predominantly positive (76\%) as a possible consequence of a hard-to-soft spectral evolution of the prompt spectrum. Lags between the LLE band (30-100 MeV) and the GBM one (10-100 keV) show a variety of behaviors: 40\% are positive, while 37\% are negative. Such negative lags may suggest the delayed emergence of an additional emission component dominating at high energies. Indeed, the spectral analysis of LLE data for 56 GRBs shows that negative lags are associated with an LLE spectral index typically harder than the high-energy power-law identified in GBM data.}
  % Conclusions heading (optional), leave it empty if necessary 
   {Spectral lags of LLE data can be exploited as a diagnostic tool to identify and characterize emission components in GRBs, highlighting the importance of combining temporal and spectral analyses to advance our understanding of GRB emission mechanisms.}

   \maketitle
%
%-------------------------------------------------------------------

\section{Introduction}\label{Introduction}

% What are GRBs and  what are spectral lags 
GRBs are observed as transient events of short duration in the keV–MeV energy range probed from space. They signpost cataclysmic phenomena such as the explosions of massive stars or the mergers of compact objects in binary systems (e.g. \citealt{piran2004physics}). GRBs are categorized based on their observed duration into two types: short GRBs ($<$2 sec), typically linked to compact object mergers, and long GRBs ($>$2 sec), generally associated with the collapse of massive stars\footnote{This apparent division has been recently put into question by the discovery of long duration GRBs (e.g. 211211A - \citealt{zhong2023grb}, 230307A - \citealt{bulla2023grb}) with a compact binary progenitor and vice versa (e.g. 200826A - \citealt{rossi2022peculiar}).}. The nature of the prompt emission of GRBs is still unclear. Over the keV-MeV energy range, during the prompt phase, different emission mechanisms may concur to produce the observed signal and its temporal and spectral signatures. 

% Positive, Negative and transitions of spectral lags 
One observable fundamental to  understand the origin of prompt emission is the spectral lag, i.e. the time delay between photons of different energies (\citealt{1995A&A...300..746C}). Spectral lags are found to be a common feature of long GRBs (\citealt{norris2000connection, ukwatta2010spectral, norris2006short}) despite with non universal features. By analysing almost 2000 GRBs detected by the Burst And Transient Source Experiment (BATSE) on board the Compton Gamma Ray Observatory (CGRO), \cite{hakkila2007gamma} found mostly positive lags, when the emission in the soft energy bands (namely 20–50 keV, 50–100 keV) anticipates that in hard energy bands (namely, 100–300 keV and 300 keV–2 MeV). Only $\sim$ 15\% of the events showed negative lags, i.e. high-energy photons lagging behind low-energy ones. With a sample of 56 GRBs detected by the Burst Alert Telescope (BAT) on board the Neil Gehrels {\it Swift} satellite, \cite{bernardini2015comparing} found that half of the long GRBs have a positive lag and half a lag consistent with zero, based on spectral lag measurements in the 100–150 keV and 200–250 keV rest-frame energy bands. The latter feature is also typical of short GRBs (\citealt{yi2006spectral, norris2006short, bernardini2015comparing, tsvetkova2017konus, lysenko2024third}). Moreover, some GRBs exhibit a transition from positive to negative spectral lags within their prompt emission duration. For instance, GRB 160625B (\citealt{wei2017new, gunapati2022variational}) showed a clear change in the lag behavior around $\sim$ 8 MeV, where the lag transitions from positive to negative when comparing soft and hard energy bands, which was interpreted as a possible Quantum Gravity effect. \cite{liang2023spectral} attributes such a transition to spectral evolution effects arising from the emission dynamics. 

% Entity of spectral lags and comparison with GRB variability 
Spectral lags have been typically computed by cross-correlating GRB light curves over different energy bands, distributed in the keV-MeV energy range, sampling the prompt emission (\citealt{norris2000connection, li2004multiwavelength, li2012spectral, chen2005distribution, ukwatta2010spectral,band1997gamma}). 
%by means of the cross-correlation function (CCF) technique (\citealt{band1997gamma}). 
Typical values of the spectral lags range from $\sim$ milliseconds to several seconds (\citealt{li2012spectral}).  Spectral lags have been measured also in GRBs with prominent X-ray flares. \cite{chang2021comprehensive} correlated the light curves of X-ray flares (in the 0.3-1.5 keV and 1.5-10 keV energy bands) reporting that 89.8\% of the flares within a sample of 48 multiflare GRBs observed by the X-Ray Telescope (XRT) onboard {\it Swift}
displayed positive lags  while 9.5\% exhibited negative lags. In the GeV energy range, sampled by the Fermi Large Area Telescope (LAT), negative lags have been measured in individual GRBs (\citealt{ackermann2013multiwavelength, ajello2019decade, 2019ICRC...36..555B}) or through a systematic analysis of bright GRBs detected by \emph{Fermi}/LAT (\citealt{castignani2014time}).

In addition to a near-zero lag,  short GRBs exhibit  shorter minimum variability timescales ($\sim$0.024 s on average) compared to long GRBs ($\sim$0.25 s) as reported by \cite{maclachlan2013minimum}. Both temporal properties, namely spectral lags and variability, underscore possible intrinsic differences in the emission mechanism of short GRBs compared to long ones, as further supported by distinctive prompt emission properties  (\citealt{Ghirlanda2004,Ghirlanda2009}).

However, while the variability timescale has been estimated in different ways in the literature, spectral lags are always determined in a consistent and standardized manner, namely through the Cross Correlation Function (CCF). Therefore, they provide a "relative" measurement that can be reliably compared across different bursts.

Spectral lags can provide  insights into the nature of the emission mechanism(s) of GRBs. For instance,  positive spectral lags (i.e., high energy photons preceding low energy ones, by definition) may be the result of the spectral evolution towards low energies (softening)  of a single emission component during the prompt phase of the burst (\citealt{ryde2005interpretations}). Analytical models further suggest that positive spectral lags in the framework of the prompt emission are linked to  the temporal evolution of some parameters of GRBs such as their peak energy, hardness-ratio, low and high-energy spectral indices, coupled to the burst duration or the light curve shape (\citealt{daigne1998gamma, bocci2010lag, mochkovitch2016simple, bovsnjak2014spectral}). Alternatively, positive lags may also arise from curvature effects in which photons emitted from higher latitudes of the jet emitting surface reach the observer delayed and, owning to relativistic effects, shifted towards lower energies (\citealt{ryde2002gamma, dermer2004curvature, preece2014first, uhm2016toward}).

Negative spectral lags (high energy photons lagging low energy ones) remain more challenging to explain. Significantly negative lags may reveal the presence of an additional emission component contributing or overlapping to the prompt emission. For example, in GRB 190114C, a negative lag of $\sim$4 sec, between 10 keV and 40 MeV, signals the onset of the afterglow, driven by external shocks, which dominates as the prompt emission fades (\citealt{ravasio2019grb}). Moreover, it has been proposed that negative spectral lags may result from Inverse Compton scattering of low-energy photons by an external medium, such as a surrounding cloud near the GRB jet (\citealt{chakrabarti2018spectral}). More recently, \cite{vyas2021backscattering} demonstrated that positive lags can naturally arise from the curvature of a backscattering photosphere in an expanding GRB jet. Moreover, the jet structure may in principle produce positive and negative spectral lags (\citealt{vyas2024unified}). 

However, spectral and temporal studies of GRBs often revealed a more complex picture, with distinct emission components contributing at different phases of the burst. Claims regarding the presence of an additional emission component are common, but there is limited consistency among the findings. Certain GRBs exhibit spectral features that evolve over specific time intervals, which potentially originate from either thermal processes (\citealt{ghirlanda2003extremely, ryde2004cooling, zhang2011comprehensive}) or synchrotron-like mechanisms (\citealt{meszaros1994delayed, daigne1998gamma, bovsnjak2009prompt}). In some cases, GRB spectra remain soft even after the apparent end of the prompt emission, marking the transition to the afterglow phase (\citealt{giblin1999evidence, tkachenko2000observations, ghirlanda2010onset, kumar2010external}).

%Our work: what's new and why is important to examine LLE emission
So far, spectral lags in GRBs have primarily been measured by considering data from the same detector (e.g. the GBM/\emph{Fermi}, BATSE/CGRO, BAT/\emph{Swift}, XRT/\emph{Swift}) and by cross-correlation analysis of the light curves in fairly close energy ranges. \cite{castignani2014time} extended this methodology by cross-correlating light curves from the two \emph{Fermi} detectors, LAT (100 MeV - 300 GeV) and GBM (8 keV - 1 MeV), for five bright GRBs.   

\begin{figure}[!ht]
    \centering
    \includegraphics[width=0.5\textwidth]{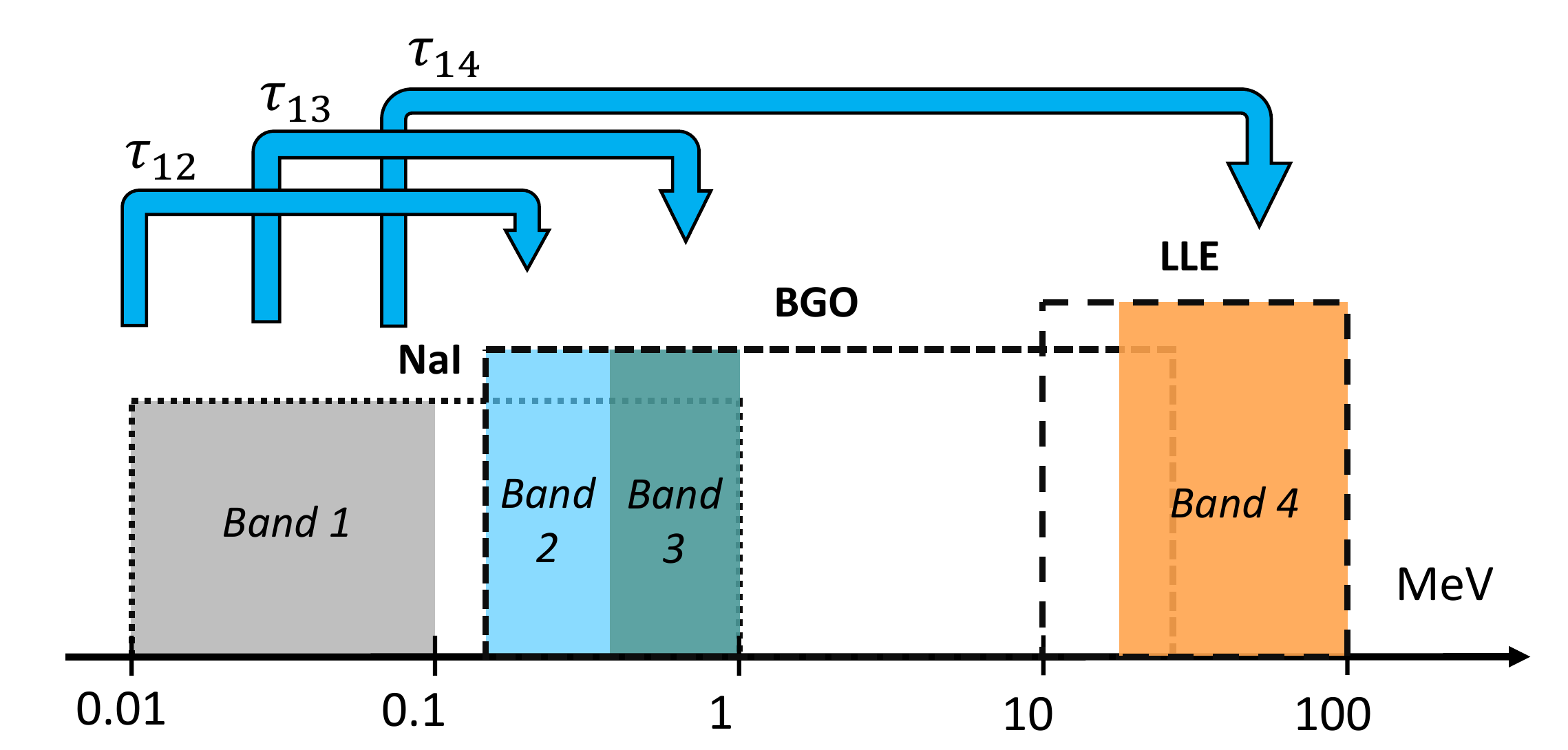} 
    \caption{Representation of the four energy ranges considered for the computation of spectral lags (shaded filled regions). Blocks defined by long and short dashed and  dotted lines represent the full energy range of each \emph{Fermi} detector (as labelled). For each GRB in the sample, the light curve extracted in Band 1 (10 - 100 keV) is cross-correlated (blue arrows) with those obtained in Band 2 (150 - 500 keV), Band 3 (500 keV - 1 MeV) and Band 4 (30 - 100 MeV) in order to obtain the corresponding lags $\tau_{12}$, $\tau_{13}$ and $\tau_{14}$.}
    \label{schematic}
\end{figure}

In this work, we measure the spectral lags of 70 GRBs detected by \emph{Fermi} by cross--correlating (see Fig.\ref{schematic}) the light curves in the 10 - 100 keV bands of GBM-NaI detectors with the light curves in two GBM-BGO bands (within 150 keV and 1 MeV) and with the light curve of the \emph{Fermi}-LAT Low Energy (LLE) events (30 - 100 MeV). 
This work is motivated by the possibility of observing, in the high energy LLE data, the compresence of both the prompt and the early-afterglow emission of external origin. Our goal is thus to explore the potential of spectral lags as a diagnostic tool for identifying distinct spectral components or emission processes in GRBs, emphasizing their utility in probing the physical mechanisms driving GRB emission.
%This study represents the first systematic use of LLE data for studying the temporal properties of GRBs. 

% Methods 
%We provide time lags for 70 \emph{Fermi}-GRBs between the 10 - 100 keV range (from GBM) and three energy ranges with increasing energy, reaching up to 100 MeV. To perform the cross-correlation analysis and obtain time lags, the Discrete Correlation Function (DCF) Method and Monte Carlo (MC) simulations were adopted. For 56 GRBs in the sample, a spectral analysis of their LLE data is also performed to obtain the respective photon indices. 

%Overview of the sections
We present the selected sample in Section \ref{sample_reduction}. The method used to estimate the lags and the LLE spectral analysis procedure are described in Sections \ref{Methods} and \ref{Spectral_analysis_LLE}, respectively. The results of our analysis are presented in Section \ref{Results} and their comparison with the GRB prompt emission properties in Section \ref{prompt_emission}. In Section \ref{Spectral Results} we compare spectral lags with the results of the spectral analysis of the LLE data. We summarize the main findings in Section \ref{Conclusions}.

%--------------------------------------------------------------------

\section{Sample Selection and Data Reduction} \label{sample_reduction}

% Brief introduction on Fermi and its detectors 
Since its launch into orbit on June 11, 2008, the \emph{Fermi} Gamma-ray Space Telescope has enabled to study GRB properties across a broad energy range. This capability is achieved through its two scientific instruments, the GBM (\citealt{meegan2009fermi}) and the LAT (\citealt{atwood2009large}). GBM consists of 12 sodium iodide (NaI) detectors and two bismuth germanate (BGO) detectors, which are sensitive to energy ranges of 8 keV -- 1 MeV and 150 keV -- 40 MeV, respectively. The LAT is a pair production telescope, detecting $\gamma$-rays from $\sim$10 MeV to over 300 GeV. In particular, the LAT Low Energy (LLE) technique is a specialized analysis method designed to investigate bright transient events, such as GRBs and solar flares, within the $\sim$ 10 MeV- 100 MeV energy range (see, e.g. \citealt{2010arXiv1002.2617P}).

At the time of writing, GBM has detected nearly 4000 GRBs. However, only 78  bursts have corresponding LLE data available for analysis, as reported in the \emph{Fermi}-LLE Catalog (FERMILLE\footnote{\url{https://heasarc.gsfc.nasa.gov/W3Browse/fermi/fermille.html}}). In our analysis, we considered all GRBs available in both the LLE and GBM catalogs\footnote{\url{https://heasarc.gsfc.nasa.gov/W3Browse/fermi/fermigbrst.html}}. 
 For 70 GRBs we inferred a robust estimate of the spectral lags. Seven GRBs exhibit low-statistics LLE data, thus preventing us from an accurate estimate of their lags through the methods described in Section \ref{Methods}. The bright GRB 130427A, despite having a LLE signal detection significance of $15\sigma$, was excluded from the analysis as its GBM light curve suffers from saturation effects. 

On average, a reliable estimate of the spectral lags is obtained for 70 GRBs that have a detection significance in the LLE energy range of $\ge 8\sigma$  (Table 2 in \citealt{ajello2019decade}). 

\subsection{Fermi-GBM data reduction} \label{GBM_reduction}

GBM data were acquired from the \emph{Fermi}-GBM Burst Catalog and processed with the \texttt{GBM Data Tools} Python package\footnote{\url{https://fermi.gsfc.nasa.gov/ssc/data/analysis/gbm/gbm_data_tools/gdt-docs/}}. For each GRB in our sample, we considered data from the NaI and BGO detectors that display the strongest signal, as identified from the quick-look light curves available in the catalog (\citealt{von2020fourth}). Data covering the time window between 50 seconds before and 300 seconds after the trigger time were selected. Each NaI event list was filtered in order to include only photons with energy between 10 keV and 100 keV  (Band 1 - grey shaded region in Fig.~\ref{schematic}). The event list of the corresponding BGO detector was filtered to determine two distinct energy ranges: the first energy range contains photons with energies between 150 keV and 500 keV (Band 2 - blue region in Fig.~\ref{schematic}); the second energy range (Band 3 - green region in Fig.~\ref{schematic}), considers photons with energies between 500 keV and 1 MeV. For each GRB, light curves obtained in Bands 1, 2 and 3 are then extracted by binning the filtered event lists in time bins of 0.1 s for long GRBs and 0.01 s for short GRBs. 

\begin{figure}[h!]
    \centering
    \includegraphics[width=0.5\textwidth]{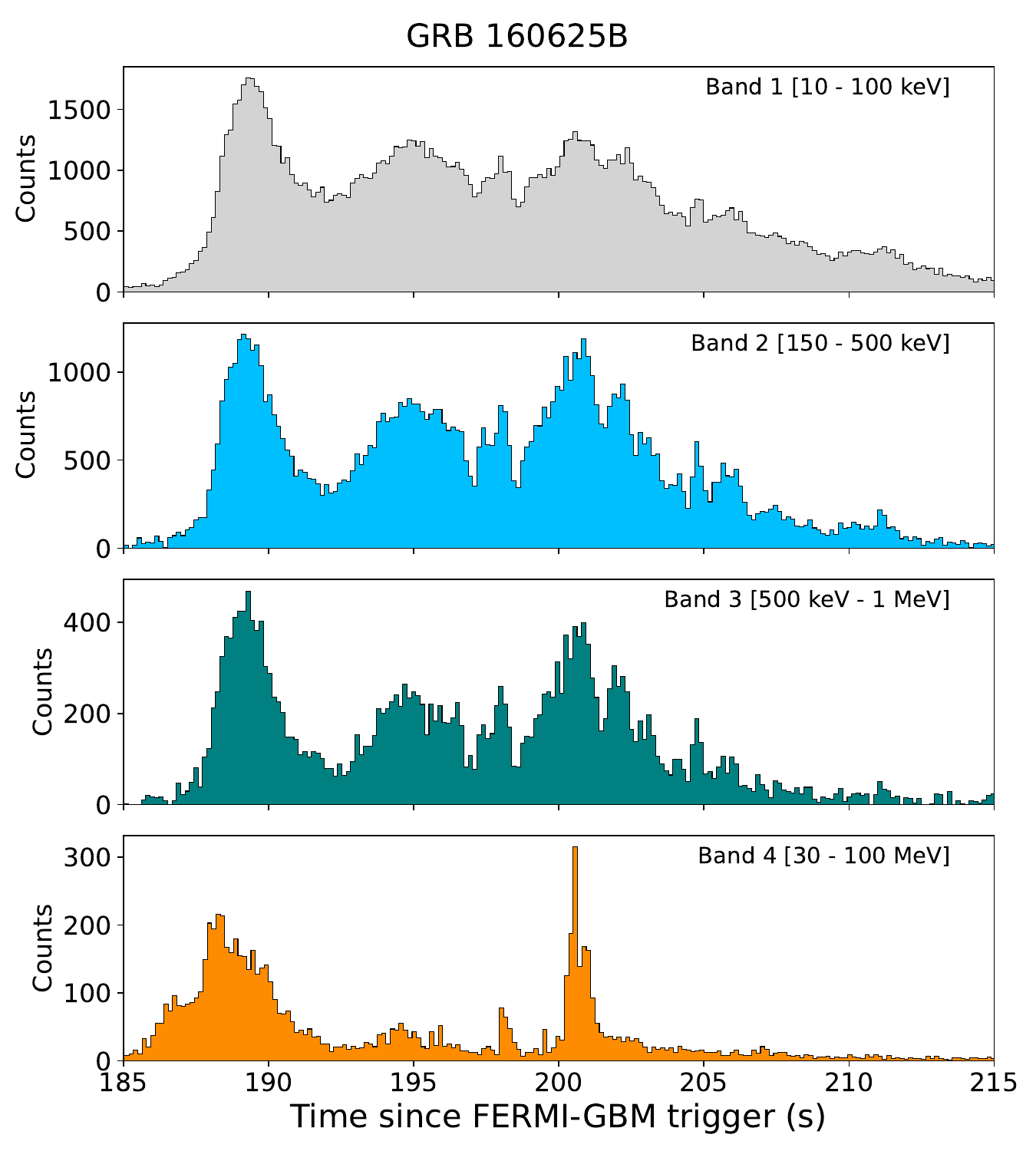} 
    \caption{Light curves of GRB 160625B. From top to bottom: Band 1 (10 - 100 keV), Band 2 (150 keV - 500 keV), Band 3 (500 keV - 1 MeV), and Band 4 (30 MeV - 100 MeV) background-subtracted light curves, binned at 0.1 s. The time interval adopted for the cross-correlation is [185.0, 215.0] s, which accounts for the presence of a burst precursor, with the main emission episode being delayed by $\sim$ 185 sec with respect to the trigger time.}
    \label{light_curves}
\end{figure}

As an example, the light curves of GRB~160625B extracted in Band 1, 2 and 3 are shown in the top three panels of Fig.~\ref{light_curves}. 

\subsection{Fermi-LLE data reduction} \label{LLE_reduction}

LLE data of each GRB were obtained from the FERMILLE Online Catalog. Events were pre-selected using the method described in \cite{2010arXiv1002.2617P}. Subsequently, we refined the selection to include only events occurring from 500 seconds prior to 1000 seconds after the LAT trigger time (available in FERMILLE), and restricted to photons with energies between 30 MeV and 100 MeV (hereafter referred to as Band 4—the orange region in Fig.~\ref{schematic}). These events were grouped into time bins of 0.1 s for long GRBs and 0.01 s for short GRBs to generate the corresponding light curves.

The list of selected energy bands adopted for the analysis is reported in Table \ref{energy_ranges}. The choice of these particular four bands  is driven by the need to have separate, non-overlapping ranges, while also ensuring sufficient statistics. Additionally, multiple energy ranges allow to study the evolution of time lags with energy, as performed in Section \ref{prompt_emission}.  

\subsection{Background Subtraction} \label{background} 

Both GBM (Band 1, 2 and 3) and LLE (Band 4) light curves are processed in a consistent manner for background subtraction. For each light curve, we subtract the time-dependent background by selecting a time interval before and one after the burst emission episode. We then fit polynomials of degrees 0 to 4 to the union of these two intervals and determine the best-fit polynomial by means of the least $\chi^2$ method. We then interpolate the best-fit polynomial over the time interval during which the burst occurs in order to subtract the background.

\begin{table}[]
\centering
\begin{tabular}{l l l}
\hline
Band   & Instrument & Energy (MeV) \\ \hline
Band 1 & GBM-NaI    & 0.01 - 0.1   \\ 
Band 2 & GBM-BGO    & 0.15 - 0.5   \\ 
Band 3 & GBM-BGO    & 0.5 - 1      \\ 
Band 4 & LAT (LLE)  & 30 - 100     \\ \hline
\end{tabular}
\caption{Energy bands (reported in MeV) adopted for the cross-correlation analysis of our sample of GRBs, in MeV. For each GRB, light curve extracted in Band 1, which considers GBM-NaI events, is cross-correlated to those of Band 2, Band 3 (from BGO detector) and Band 4 (LAT-LLE events).}
\label{energy_ranges}
\end{table}

\begin{figure*}[!ht]
    \centering
    \includegraphics[width=0.9\textwidth]{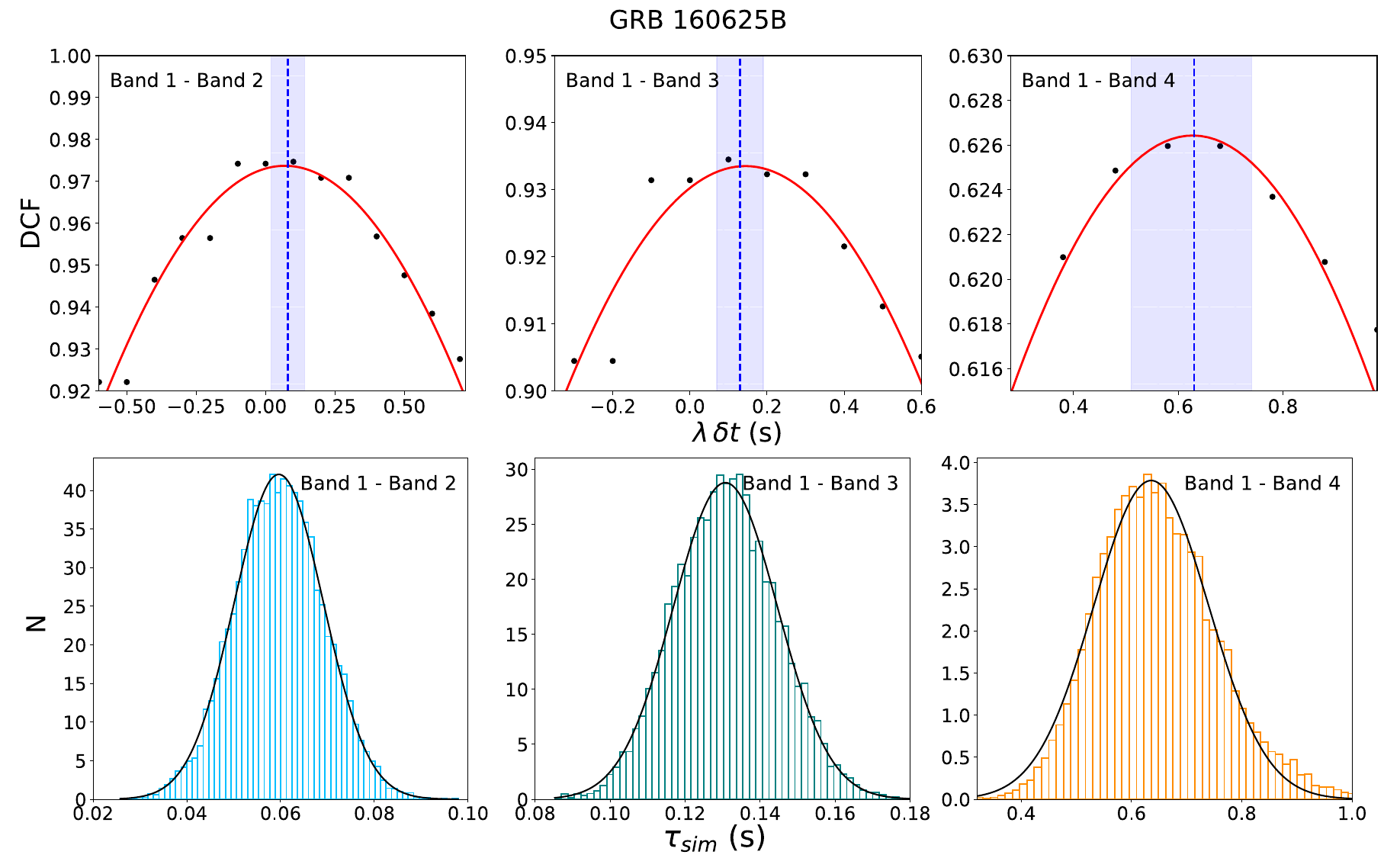} 
    \caption{Lag computation for GRB 160625B. Upper panel: DCF as a function of the time delay (in seconds - bins of 0.1 s) between Band 1 (10 - 100 keV) background-subtracted light curve and those of Band 2 (150 - 500 keV) (left plot), Band 3 (500 keV - 1 MeV) (central plot) and Band 4 (30 - 100 MeV) (right plot). In each plot the red line represents the best fit to the DCF with an asymmetric Gaussian model. Blue-dashed lines mark the maximum of the asymmetric Gaussian model, while coloured areas mark its 1$\sigma$ uncertainty. %DCFs are obtained by adopting a time bin $\delta t =$0.1 s. 
    The latter is estimated through Monte Carlo simulations of GRB 160625B time lags (bottom panels). From left to right: distribution of 10 000 synthetic $\tau_{12}, \tau_{13}$ and $\tau_{14}$ spectral lags obtained through a flux-randomization method. The black lines show the best fit with a symmetric Gaussian. The mean value and the standard deviation of each Gaussian represent the best value of the spectral lag and its corresponding 1$\sigma$ uncertainty.}
    \label{dcf_and_montecarlo}
\end{figure*}

%--------------------------------------------------------------------

\section {Time Lag Calculation} \label{Methods}

\subsection{Discrete Correlation Function Method} \label{DCF_section}

In this work we want to evaluate spectral lags between the selected GBM-NaI energy range ( Band 1 -- 10 - 100 keV) at the lowest energies, and those of GBM-BGOs (150 - 500 keV and 500 keV - 1 MeV, i.e. Band 2 and 3) and LAT-LLE (Band 4 -- 30 - 100 MeV) respectively. We will thus refer to the corresponding spectral lags as $\tau_{12}\,,\,\tau_{13}$ and $\tau_{14}$ (as highlighted in Fig.~\ref{schematic}). In order to evaluate the temporal correlation between the background-subtracted light curve in Band 1 and those of Band 2, Band 3 and Band 4, we employ the discrete Cross Correlation Function (DCF) method. Unlike the traditional method, we also adopted a non-mean-subtracted version of the DCF as proposed by \cite{band1997gamma}, which is better suited for analyzing transient phenomena like GRBs (as already adopted in \citealt{bernardini2015comparing}). The DCF can be expressed as: 

\begin{equation} \label{DCF}
\text{DCF}_{x, y}(\lambda \delta t) = \frac{\sum _{i = 1}^{N - \lambda} x_{i} \,\, \,y_{(i + \lambda)}}{\sqrt{\sum_i x_i^2 \sum_i y_i^2}}\,\,,
\end{equation}

where, $\lambda \delta t$ represents the time lag, with $\lambda = (..., -1, 0, 1, ...)$ being an integer, and $\delta t$ being the time bin of the DCF. $N$ is the total number of data points in the light curve, $x_i$ and $y_i$ refer to the count rates in Band 1, and in Band 2, Band 3 or Band 4 i-th energy channels, respectively. For each GRB, the time interval in which the DCF is evaluated corresponds to the one which contains the primary emission episodes in all four energy bands (see Table \ref{table_lags}). This selection allows to avoid DCF linking of unrelated structures.

For each pair of light curves (i.e., those in Band 1 and Band 2, 3 or 4) we compute the DCF across a range of $\lambda \delta t$. The spectral lag $\tau$ is then defined as the time delay for which we have the global maximum of the DCF:

\begin{equation}\label {time_lag}
\text{DCF}_{x, y} (\tau) = \max \left[\,\, \text{DCF}_{x, y} (\lambda \delta t)\,\,\right]\,\,,
\end{equation}

To identify $\tau$, we fitted an asymmetric Gaussian model to the DCF as a function of the time delay, as already performed in e.g. \cite{castignani2014time} and \cite{bernardini2015comparing}. 

\begin{equation}
    \text{DCF}_{xy}(t) = K +
\begin{cases} 
A \exp\left(-\frac{(t - \tau)^2}{2\Sigma_l^2}\right), & \,\, t \leq \tau, \\
A \exp\left(-\frac{(t - \tau)^2}{2\Sigma_r^2}\right), & \,\, t > \tau.
\end{cases}
\end{equation}

The fitting process was carried out using the least $\chi^2$ method. It is worth noticing that fitting a continuous function to the DCF enables us to obtain values of $\tau$  (corresponding to DCF maxima) that can be smaller than the time resolution $\delta t$ of the light curves. 

As already emphasized by \cite{band1997gamma}, DCFs are often asymmetric as a direct result of the asymmetric nature of GRB pulses, so that an asymmetric Gaussian model strikes an optimal choice for fitting. In Fig.~\ref{dcf_and_montecarlo} (upper panel), we present the lag computation of GRB 160625B as an example, for the three cases of $\tau_{12}$,  $\tau_{13}$ and  $\tau_{14}$. The red solid lines show the asymmetric gaussian fit around the peak of each DCF. Vertical blue-dashed lines identify the position of the time lag (the maximum of the DCF), while blue shaded regions represent the respective uncertainty of each time lag at 1$\sigma$. Notably, the number of points of the peak decreases when transitioning from the DCF obtained between Band 1 and Band 2 and 3, to the DCF between Band 1 and 4, as a consequence of the decreasing statistic at higher energies, which lead to greater scattering in the DCF.

For highly complex light curves of some GRBs which can include multiple emission episodes, more or less distinct from one another, the DCF may exhibit a combination of multiple similar-height peaks, rather than a single well-defined one. In such cases determining a single global maximum is challenging. In our analysis, we consider two or more DCF peaks as comparable if the difference between their heights is less than 10\% of the value of the largest peak. In such cases, instead of selecting a single time lag, we take into account all the time lags corresponding to these comparable peaks.

\subsection{Time Lag Error Estimation} \label{MC}

We compute the best value of $\tau$ and its corresponding uncertainty by applying a flux-randomization method prescribed in \cite{peterson1998uncertainties}; we generate $N = 10 000$ realizations for each light curve extracted in Band 1, 2, 3 and 4 based on their count rate at each $i$-th time bin, $\bar{x_i}$, and its error $\Delta x_i$ (Poissonian error). Then, for all cases, we subtract the background for each simulated light curve in the same manner described in Section \ref{background}. We then compute the DCFs for each pair of simulated light curves and fit them with an asymmetric Gaussian, thus obtaining $N$ synthetic values of $\tau$. 

Given that $N \gg 1$, the central limit theorem suggests that the distribution of synthetic $\tau$ should approximate a Gaussian. Accordingly, we fit each distribution with a Gaussian function and take its mean and standard deviation as the best value of $\tau$ and its  $1\sigma$ uncertainty, respectively. This approach yields time lag distributions of each GRB, allowing for an assessment of  $\tau$ and its uncertainty which does not depend on the function used for fitting the DCF peak (\citealt{ukwatta2010spectral}). In Fig.~\ref{dcf_and_montecarlo} (lower panel) the spectral lags distribution of the Monte Carlo simulation described above are shown for the case of  GRB~160625A, as an example. As for the estimation of the best value of the spectral lag, this methodology allow us to obtain uncertainties that can be smaller than the resolution $\delta t$ of the light curves. 

The accuracy and precision of time lag estimation are influenced by the signal-to-noise ratio (S/N) of the light curves (\citealt{band1997gamma, ukwatta2010spectral}). In cases of noisy light curves, the DCF becomes more scattered, and the maximum of the function decreases, resulting in a less accurate determination of $\tau$. This led to the exclusion, in our analysis, of seven GRBs due to their low LLE signals. As previously emphasized, using a continuous function to fit the DCF enables the estimation of spectral lags that can be smaller than the time binning $\delta t$, ensuring no significant bias is introduced by varying bin sizes. However, the selection of the time bin is driven by the need for accurate sampling of the DCF with at least 5 points and for having an adequate signal-to-noise ratio across all time bins. Therefore, for long GRBs, we choose a DCF time bin of 0.1 s, and for the four short GRBs in our sample, we use a bin of 0.01 s, matching the bin size of the extracted light curves.

\section{LLE Spectral Analysis} \label{Spectral_analysis_LLE}

For every GRB in our sample, we perform a time-integrated spectral analysis using LLE data within the 30–100 MeV energy band. LLE spectral data files, along with their response matrix files (rsp2), are retrieved and produced with the public software \texttt{GTBURST}. This software allows us to extract the LLE spectrum for each GRB over a specified time interval starting from its LLE light-curve. The background is subtracted by fitting a model to two regions of the data outside the burst emission (one before and one after the main emission, respectively). An automated fitting algorithm then determines the optimal polynomial model to represent the background. Spectral files are then retrieved, and the spectral analysis is conducted using the publicly available \texttt{XSPEC} software (version 12.13.1).

We fit each time-integrated spectrum with a power-law model with pegged normalization in the 30–100 MeV (LLE) range (\emph{pegpwrlw} model in \texttt{XSPEC}). The model is defined as:

\begin{equation} \label{power_law}
    A(E) = K E^{\,\alpha_{\text{LLE}}}\,
\end{equation}

where $\alpha_{\text{LLE}}$ is the photon index of the power law (defined as negative) and $K$ is the normalization, representing the model flux in units of $10^{-12}\,\text{erg/cm}^2/\text{s}$ integrated over the 30 - 100 MeV energy range. While it is known that some bursts may exhibit a high-energy cutoff around a few hundred MeV (\citealt{ravasio2024insights}), our focus in this study is on assessing the consistency of the LLE spectrum at lower energies with the prompt emission spectrum as measured by GBM, particularly by the BGO detectors. Consequently, we do not account for potential spectral features such as high-energy cutoffs or spectral break. For 7 GRBs in our sample (namely GRBs 090227B, 120226A, 130504C, 131216A, 140102A, 150403A and 150523A), it is necessary to extend the energy range down to 10 MeV in order to obtain reliable estimates of the spectral parameters. As reported in the Fermi Caveats for LLE data\footnote{\url{https://fermi.gsfc.nasa.gov/ssc/data/analysis/LAT_caveats_p8r2.html}}, the effect of energy dispersion can be significant at low energies, and spectral analysis below 30 MeV is generally discouraged due to poor event reconstruction. However, for these GRBs, we incorporate data between 10 and 30 MeV to increase the statistical significance of the fit, which would not have been achievable by limiting the analysis to the 30–100 MeV range. Moreover, we verified that extending the spectral data down to 10 MeV allows us to reliably calculate the errors on the spectral parameters, which were otherwise unconstrained when fitting the data only in the 30-100 MeV range. We verified that the central values of the  parameters obtained in this extended range (10-100 MeV) are consistent with those derived from the narrower range, confirming that the inclusion of data down to 10 MeV does not introduce any systematic effect. 

\begin{figure*}[!ht]
    \centering
    \includegraphics[width=0.9\textwidth]{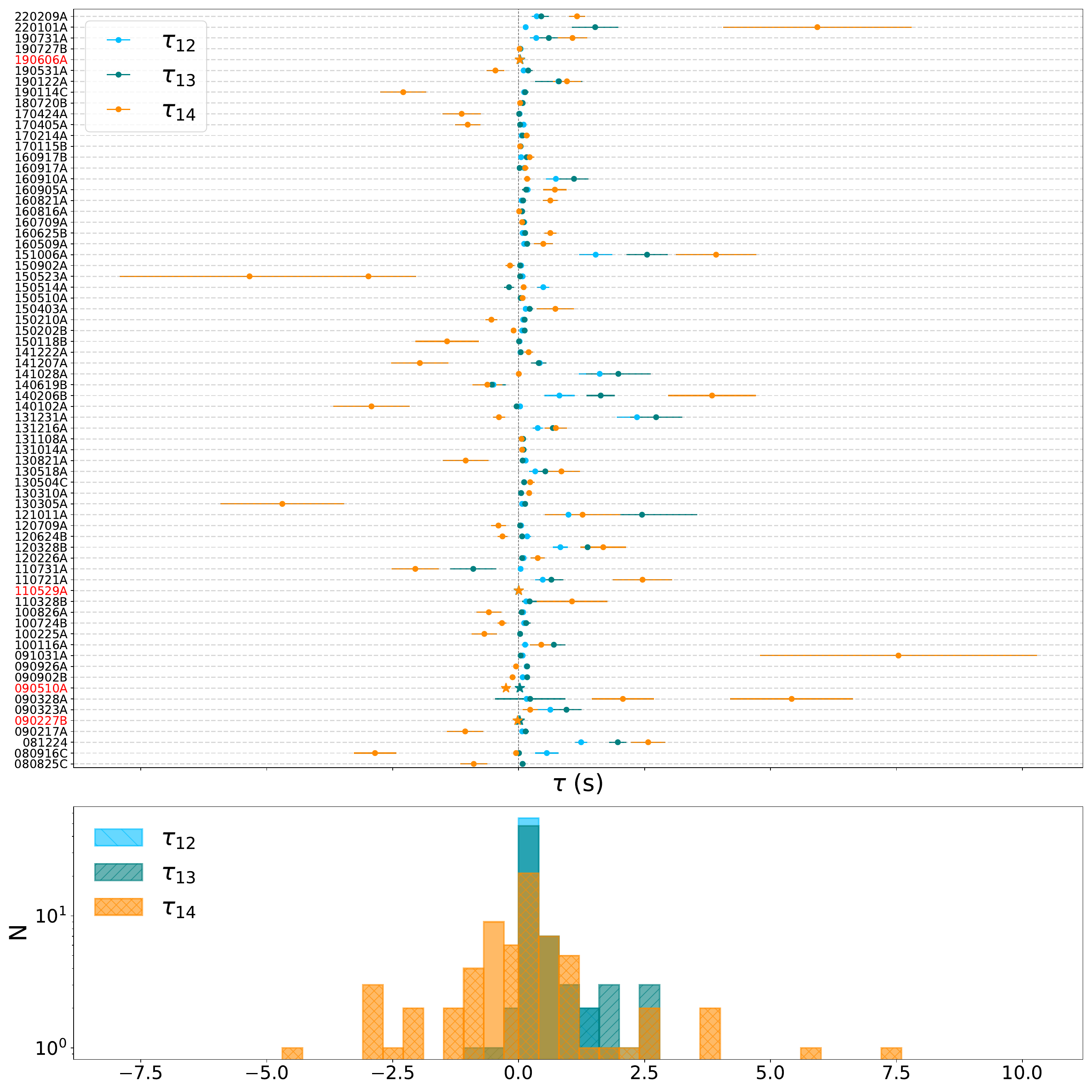} 
    \caption{Spectral lags for the 70 GRBs in our sample. Upper panel: lags between Band 1 and Band 2 ($\tau_{12}$, blue points), Band 3 ($\tau_{13}$, green points) and Band 4 ($\tau_{14}$, orange points) and their 1$\sigma$ uncertainties  (listed in Table~\ref{table_lags}). Short GRBs are marked with star symbols and their names are in red color. Spectral lags $\tau$ values are represented  in symmetric-logarithmic scale for visualization purposes. Lower panel: distributions of spectral lag values.}
    \label{all_lags}
\end{figure*}

\begin{figure*}[!ht]
    \centering
    \includegraphics[width=0.85\textwidth]{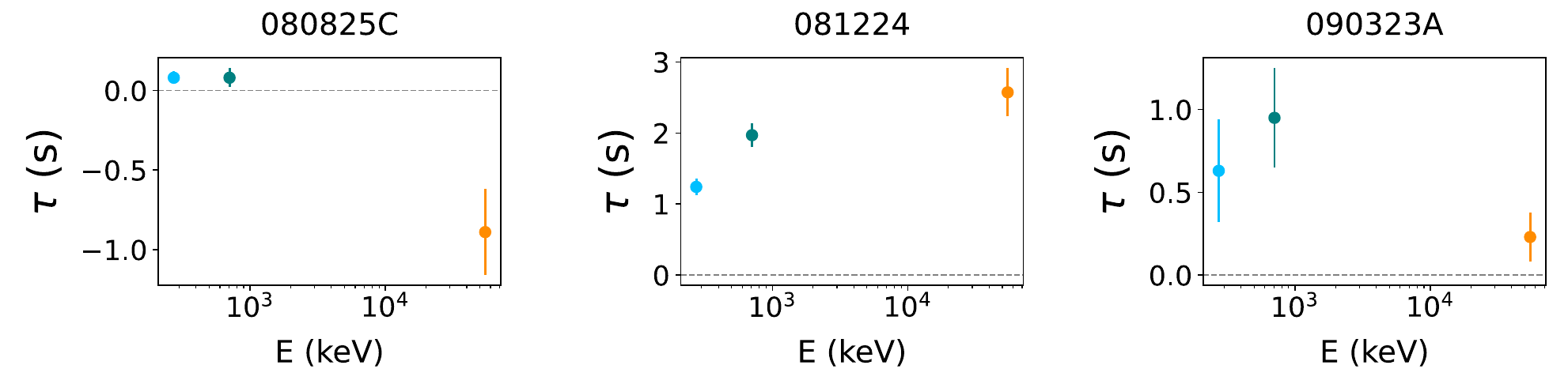} 
    \caption{$\tau_{12}$ (blue point) $\tau_{13}$ (green point) and $\tau_{14}$ (orange point) lags as a function of the mean energy of the respective channel. Each plot represents an example of GRB which show one characteristic trend of the time lag versus energy. From left to right: transition from positive values of $\tau_{12}$ and $\tau_{13}$ to negative $\tau_{14}$ values, increasing time lag with energy and an example of a more hybrid, scattered trend.} 
    \label{example_energy}
\end{figure*}

\section{Results} \label{Results}

\subsection{Spectral Lag Behavior} \label{lags_behaviour}

Fig.~\ref{all_lags} (upper panel) shows the spectral lags $\tau_{12}$ (blue points), $\tau_{13}$ (green points) and $\tau_{14}$ (orange points), along with their corresponding 1$\sigma$ uncertainties, for all the 70 GRBs in our sample (in chronological order, from bottom to top). GRB names marked in red correspond to the four short bursts in the sample (with their spectral lags indicated by stars in the figure). All spectral lags and their uncertainties are also reported in Table~\ref{table_lags}, for each GRB. 

Considering a significance threshold of $\geq 2\sigma$, we find that 40\% of $\tau_{\text{14}}$ lags are significantly positive, while 37\% are significantly negative. For $\tau_{\text{12}}$ and $\tau_{\text{13}}$, approximately 76\% and 61\% of the delays are significantly positive, respectively, with no significantly (at the $\geq 2\sigma$ confidence level) negative values. Overall, spectral lags are found to range from fractions of a second to a few seconds. However, over our entire sample of GRBs, we can observe different behaviors by considering both the distinction between short and long GRBs and the different energy channels in which the spectral lags are computed, suggesting that these delays are related to GRB physics and not to purely instrumental or statistical effects.

\subsection{Spectral Lags of Long GRBs} \label{lags_long_grb}

For long GRBs, the vast majority of spectral lags $\tau_{\text{12}}$ and $\tau_{\text{13}}$ values are  positive (98.6\% and 94.3\% for Band 2 and Band 3, respectively), meaning that soft photons systematically lag behind hard ones. Moreover, the few instances of negative $\tau_{12}$ and $\tau_{13}$ are not significantly different from 0 at the $\geq 2\sigma$ confidence level (see Table~\ref{table_lags}), suggesting that the predominance of positive delays is a robust feature in our sample. Values of $\tau_{12}$ and  $\tau_{13}$ in the sample are on average on the order of fractions of a second, with the largest spectral lags obtained for GRBs 131231A (for which $\tau_{12} \sim2.4$s and $\tau_{13} \sim 2.7$ s), and GRB 151006A (with $\tau_{13} \sim2.6$ s). On the contrary, $\tau_{14}$ values for long GRBs are almost equally divided between positive (60\%) and negative (40\%) and significantly longer, reaching up to several seconds (see Fig.~\ref{all_lags}). In this case, the largest time lags are observed for GRBs 091031A (with $\tau_{14} \sim7.5$ s) and 130305A ($\tau_{14} \sim-4.7$ s).

The differences in sign and magnitude we observe between values of $\tau_{12}$, $\tau_{13}$, and $\tau_{14}$ lead to two markedly distinct distributions of these lags, as shown in Fig.~\ref{all_lags} (lower panel). Specifically, the distributions of $\tau_{12}$ and $\tau_{13}$ spectral lags (blue and green-shaded regions in the plot, respectively) are highly asymmetric and concentrated at positive values. In contrast, the distribution of $\tau_{14}$ lags (orange-shaded region) is nearly symmetric around zero and extends to larger absolute values. 

Moreover, in GRB~080916C, 090328A and 150523A,  a second value of $\tau_{14}$ is inferred from their DCF, specifically at $(-0.05 \pm 0.02)$s, $(5.42\pm 1.22)$s, and $(-5.34 \pm 2.58)$s, respectively (see Fig.\ref{all_lags}). Their primary values of $\tau_{14}$, on the other hand, are listed in Table \ref{table_lags}, according to all other GRBs in the sample. The presence of such secondary time lags occurs as the corresponding DCFs display two distinct peaks with comparable heigh, according to the criteria described in Sec.~\ref{Methods}. Interestingly, all DCFs computed between Band 1 and Bands 2 and 3 show a distinct single global maximum, so that only a single value of $\tau_{12}$ and $\tau_{13}$ is given, for all GRBs in the sample.\\

\subsection{Spectral Lags of Short GRBs} \label{lags_short_grb}

The short GRBs 090227B, 110529A and 190606A  exhibit lag values that are consistent with zero in all cases (at the $\geq 2\sigma$ confidence level, see Table~\ref{table_lags}). This result aligns with earlier findings for this class of bursts (\citealt{yi2006spectral, norris2006short, guiriec2010time, bernardini2015comparing, tsvetkova2017konus, lysenko2024third}).  However, given the small number of short GRBs in our sample, we cannot make an accurate comparison of the two populations of short and long GRBs spectral lag distributions. 

GRB 090510A is the only short GRB in the sample to show a non-negligible time lag value at the $\ge 2\sigma$ confidence level, having $\tau_{14} = -0.25 \pm 0.08$s. This is consistent with the finding of an extended and slightly delayed spectral component of afterglow origin dominating the emission from this burst in the LAT data (\citealt{ghirlanda2010onset}). 

\section{Spectral Lags and Prompt Emission Properties} \label{prompt_emission}

In this section we investigate the possible relationship between spectral lags and some characteristic properties of the prompt emission. Given that time lags are calculated relative to the GBM-NaI detector (10–100 keV), it is appropriate to compare them with other temporal and spectral properties of the prompt emission, as well as their evolution with energy, if present. 

\subsection{Energy dependence} \label{lag_vs_energy}

In Fig.~\ref{example_energy}, we show the spectral lags $\tau_{12}, \tau_{13}, \tau_{14}$ for three representative GRBs from our sample: GRB 080825C, 081224, and 09032A. These lags are plotted against the mean energy of the respective channels (Band 2, Band 3, and Band 4), defined as $E = \sqrt{E_{\text{min}}E_{\text{max}}}$, where $E_{\text{min}}$ and $E_{\text{max}}$ are the lower and upper energy bounds of each band. The corresponding plots for all GRBs in our sample are provided in Appendix \ref{appendix}.

We identify two distinct trends between the spectral lag and the mean channel energy. The first trend, illustrated by GRB~080825C in the left panel of Fig.~\ref{example_energy}, shows that the time lag decreases as the energy of the band selected to compute it, with respect to Band 1, increases. In this case, a transition occurs from positive values of $\tau_{\text{12}}$ and $\tau_{\text{13}}$ to a negative value of $\tau_{\text{14}}$. This trend is observed in 28 GRBs (40\%) in our sample, with their lag versus energy plots shown in Fig.~\ref{transition_grbs}. Notably, in some of these GRBs, $\tau_{12} > \tau_{13}$. 

A second trend, exemplified by GRB~081224 in the central panel of Fig.~\ref{example_energy}, is characterized by positive values of the three spectral lag which increases with energy. This behavior is observed in 25 out of 70 GRBs ($\sim$ 36\% of the sample), and their corresponding lag versus energy plots can be found in Fig.~\ref{positive_grbs}. 

\clearpage

\begin{table*}[]
\centering
\scalebox{0.85}{%
\begin{tabular}{l r r r r r r r r }
        \hline
          GRB name & $t_{\text{start}}$ (s) & $t_{\text{stop}}$ (s) & $\tau_{12}$ (s) & $\sigma_{12}$ (s) & $\tau_{13}$ (s) & $\sigma_{13}$ (s) & $\tau_{14}$ (s) & $\sigma_{14}$ (s) \\ \hline
          080825C & -2.0 & 40.0 & 0.08 & 0.04 & 0.08 & 0.06 & -0.89 & 0.27\\
          080916C & -2.0 & 20.0 & 0.56 & 0.23 & 0.008 & 0.006 & -2.85 & 0.42\\
          081224 & -2.0 & 15.0 & 1.24 & 0.12 & 1.97 & 0.17 & 2.57 & 0.34\\
          090217A & -5.0 & 22.0 & 0.07 & 0.04 & 0.14 & 0.07 & -1.06 & 0.36\\
          090323A & -0.5 & 80.0 & 0.63 & 0.31 & 0.95 & 0.3 & 0.23 & 0.15\\
          090328A & -0.5 & 30.0 & 0.16 & 0.08 & 0.23 & 0.7 & 2.07 & 0.62\\
          090902B & -2.0 & 25.0 & 0.08 & 0.03 & 0.17 & 0.05 & -0.12 & 0.06\\
          090926A & -1.0 & 15.0 & 0.16 & 0.05 & 0.17 & 0.05 & -0.11 & 0.04\\
          091031A & -1.0 & 25.0 & 0.08 & 0.04 & 0.04 & 0.03 & 7.54 & 2.75\\
          100116A & 80.0 & 105.0 & 0.13 & 0.07 & 0.70 & 0.23 & 0.45 & 0.23\\
          100225A & -10.0 & 20.0 & 0.03 & 0.02 & 0.03 & 0.04 & -0.68 & 0.25 \\
          100724B & 5.0 & 35.0 & 0.11 & 0.05 & 0.15 & 0.09 & -0.33 & 0.09\\
          100826A & 6.0 & 37.0 & 0.09 & 0.04 & 0.06 & 0.03 & -0.59 & 0.25\\
          110328B & -0.5 & 20.0 & 0.15 & 0.07 & 0.22 & 0.15 & 1.06 & 0.70\\
          110721A & -1.0 & 6.0 & 0.48 & 0.15 & 0.65 & 0.24 & 2.46 & 0.59\\
          110731A & -1.0 & 12.0 & 0.04 & 0.02 & -0.90 & 0.35 & -2.05 & 0.47\\
          120226A & -1.0 & 40.8 & 0.10 & 0.05 & 0.07 & 0.06 & 0.38 & 0.14\\
          120328B & -5.0 & 22.5 & 0.83 & 0.15 & 1.37 & 0.13 & 1.68 & 0.45\\
          120624B & -1.0 & 20.0 & 0.17 & 0.08 & 0.07 & 0.04 & -0.32 & 0.10\\
          120709A & -0.5 & 30.0 & 0.05 & 0.02 & 0.03 & 0.01 & -0.40 & 0.15\\
          121011A & -9.5 & 15.0 & 0.99 & 0.29 & 2.54 & 1.10 & 0.74 & 0.22\\
          130305A & 1.0 & 15.0 & 0.07 & 0.04 & 0.13 & 0.05 & -4.69 & 1.23\\
          130310A & 3.5 & 5.5 & 0.05 & 0.01 & 0.05 & 0.01 & 0.21 & 0.01\\
          130504C & 11.0 & 37.0 & 0.11 & 0.03 & 0.11 & 0.03 & 0.23 & 0.09\\
          130518A & 20.0 & 38.0 & 0.33 & 0.12 & 0.53 & 0.25 & 0.85 & 0.37\\
          130821A & 22.0 & 41.0 & 0.14 & 0.05 & 0.08 & 0.03 & -1.05 & 0.45\\
          131014A & 0.0 & 6.0 & 0.09 & 0.03 & 0.10 & 0.04 & 0.07 & 0.03\\
          131108A & -0.5 & 24.0 & 0.06 & 0.02 & 0.09 & 0.03 & 0.06 & 0.02\\
          131216A & -5.0 & 15.2 & 0.38 & 0.10 & 0.68 & 0.15 & 0.74 & 0.22\\
          131231A & 12.0 & 30.5 & 2.35 & 0.40 & 2.73 & 0.52 & -0.39 & 0.12\\
          140102A & -0.5 & 5.5 & 0.03 & 0.01 & -0.04 & 0.02 & -2.92 & 0.76\\
          140206B & -2.0 & 30.0 & 0.81 & 0.30 & 1.63 & 0.28 & 3.84 & 0.87\\
          140619B & -1.0 & 4.5 & -0.50 & 0.25 & -0.53 & 0.26 & -0.62 & 0.30\\
          141028A & 5.0 & 28.0 & 1.61 & 0.42 & 1.98 & 0.64 & 0.004 & 0.002\\
          141207A & -0.5 & 25.0 & 0.42 & 0.13 & 0.40 & 0.15 & -1.96 & 0.57\\
          141222A & -1.0 & 2.0 & 0.04 & 0.01 & 0.04 & 0.02 & 0.20 & 0.09\\
          150118B & -2.0 & 38.5  & 0.02 & 0.01 & 0.009 & 0.007 & -1.42 & 0.63\\
          150202B & 4.0 & 17.5 & 0.07 & 0.02 & 0.12 & 0.04 & -0.10 & 0.04\\
          150210A & -1.0 & 5.0 & 0.09 & 0.02 & 0.12 & 0.03 & -0.54 & 0.12\\
          150403A & 2.0 & 20.0 & 0.14 & 0.05 & 0.22 & 0.07 & 0.73 & 0.37\\
          150510A & -0.5 & 15.0 & 0.05 & 0.01 & 0.16 & 0.07 & 0.22 & 0.09\\
          150514A & -1.0 & 5.5 & 0.49 & 0.12 & -0.19 & 0.10 & 0.10 & 0.04\\
          150523A & 0.0 & 40.0 & 0.08 & 0.03 & 0.03 & 0.02 & -2.98 & 0.95\\
          150902A & 0.0 & 25.0 & 0.05 & 0.02 & 0.03 & 0.01 & -0.17 & 0.09\\
          151006A & -2.0 & 13.5 & 1.53 & 0.33 & 2.55 & 0.41 & 3.92 & 0.80\\
          160509A & 7.5 & 23.0 & 0.11 & 0.04 & 0.17 & 0.06 & 0.59 & 0.23\\
          160625B & 185.0 & 215.0 & 0.08 & 0.05 & 0.13 & 0.04 & 0.63 & 0.12\\
          160709A & -0.5 & 2.5 & 0.11 & 0.03 & 0.10 & 0.03 & 0.07 & 0.04\\
          160816A & -2.0 & 15.0 & 0.07 & 0.01 & 0.07 & 0.02 & 0.011 & 0.008\\
          160821A & 115.0 & 155.0 & 0.06 & 0.01 & 0.09 & 0.02 & 0.63 & 0.15\\
          160905A & 5.0 & 20.5 & 0.18 & 0.03 & 0.15 & 0.08 & 0.72 & 0.23\\
          160910A & 5.5 & 15.0 & 0.74 & 0.20 & 1.10 & 0.29 & 0.17 & 0.07\\
          160917A & -10.0 & 23.0 & 0.11 & 0.07 & 0.02 & 0.01 & 0.13 & 0.07\\
          160917B & 10.5 & 30.5 & 0.05 & 0.03 & 0.16 & 0.07 & 0.22 & 0.09\\
          170115B & -1.0 & 15.0 & 0.04 & 0.01 & 0.04 & 0.01 & 0.03 & 0.01\\
          170214A & 35.0 & 80.0 & 0.06 & 0.02 & 0.08 & 0.04 & 0.16 & 0.06\\
          170405A & 10.0 & 60.0 & 0.10 & 0.04 & 0.03 & 0.01 & -1.01 & 0.25\\
          170424A & 0.0 & 30.5 & 0.02 & 0.01 & 0.011 & 0.009 & 1.13 & 0.38\\
          180720B & 0.0 & 35.0 & 0.082 & 0.005 & 0.08 & 0.01 & 0.03 & 0.01\\
          190114C & -0.5 & 25.0 & 0.11 & 0.04 & 0.13 & 0.07 & -2.29 & 0.46\\
          190122A & -5.0 & 20.0 & 0.79 & 0.15 & 0.80 & 0.47 & 0.96 & 0.28\\
          190531B & 5.0 & 40.0 & 0.10 & 0.04 & 0.19 & 0.09 & -0.46 & 0.18\\
          190727B & 15.0 & 42.0 & 0.04 & 0.010 & 0.03 & 0.02 & 0.02 & 0.01\\
          190731A & 0.0 & 11.0 & 0.35 & 0.13 & 0.60 & 0.22 & 1.07 & 0.29\\
          220101A & 80.0 & 140.0 & 0.14 & 0.05 & 1.52 & 0.46 & 5.93 & 1.87\\
          220209A & 8.0 & 17.0 & 0.36 & 0.10 & 0.45 & 0.15 & 1.16 & 0.16\\\\

          090227B & -0.25 & 0.25 & 0.003 & 0.009 & 0.025 & 0.009 & -0.04 & 0.02\\
          090510A & -0.25 & 1.25 & 0.027 & 0.015 & 0.02 & 0.01 & -0.25 & 0.08\\
          110529A & -0.25 & 0.50 & 0.0012 & 0.0008 & 0.003 & 0.002 & 0.005 & 0.003\\
          190606A & -0.5 & 0.5 & 0.03 & 0.02 & 0.03 & 0.02 & 0.02 & 0.01\\
          \hline
\end{tabular}%
}
\caption{Spectral Lags of the 66 long (upper part) and 4 short (lower part) GRBs in our sample. GRB name, upper ($t_{\text{start}}$) and lower ($t_{\text{stop}}$) temporal boundaries adopted for applying the DCF method, spectral lags $\tau_{12}, \tau_{13}, \tau_{14}$, and lags uncertainties ($\sigma_{12}, \sigma_{13}, \sigma_{14}$).}
\label{table_lags}
\end{table*}

\clearpage

The remaining 17 GRBs (i.e. $\sim$ 24\% of the sample) exhibit more scattered $\tau$ versus $E$ trends, as seen in GRB~090323A (right panel in Fig.~\ref{example_energy}). In these cases, shown in Fig.~\ref{hybrid_grbs}, identifying a clear trend is challenging.

\subsection{$T90$ and Fluence} \label{lag_vs_t90gbm}

Fig.~\ref{t90_and_fluence} (left panel) shows  $\tau_{12}, \tau_{13}$ and $\tau_{14}$ of each GRB in our sample plotted against the duration of the bursts computed with the GBM data (i.e., the $T90_{\text{GBM}}$ parameter reported in \cite{ajello2019decade} for all LAT-GRBs detected until August, 2018). A noteworthy behavior across all spectral lags is that the largest lag absolute values are associated with intermediate values of  $T90_{\text{GBM}}\sim 10-200$ seconds. However, a more in depth analysis, which is out of the scopes of the present work, is required to determine if this relation is real or induced by the possible energy dependence of the estimate of the burst duration (\citealt{zhang2007dependence, mochkovitch2016simple}). 

Additionally, we investigated (Fig.~\ref{t90_and_fluence}) the relation between $\tau_{12}, \tau_{13}$ and $\tau_{14}$ and the GBM fluence (as given by \cite{ajello2019decade}, computed over the 10–1000 keV energy range. At lower values of the fluence, it becomes challenging to estimate significant time lags due to the low statistics of their light curves, which can results in noisier DCFs. Moreover, it can be seen that larger values of the spectral lags (both positive and negative) tend to appear at intermediate values of the fluence, within the range considered in our sample.

\section{Spectral Lags and LLE Spectra} \label{Spectral Results}

Table \ref{table_spectral} shows the results of the time-integrated spectral analysis of the LLE data for the 70 GRBs in our sample, as described in Section \ref{Spectral_analysis_LLE}. The GRB name, the time interval used for extracting the spectrum, the best fit parameters (the photon index, $\alpha_{\text{LLE}}$ and the normalization), and the value of the PG-Statistic over the degrees of freedom are reported. Errors on the spectral parameters represent the 90\% confidence level. We could reliably fit the spectra of 56 out of 70 GRBs in our sample. For 14 GRBs in our sample it was not possible to constrain the spectral parameters due to insufficient statistics of their LLE spectra.

Spectral evolution of GRBs can be inherently linked to their time lags evaluated in different energy bands. Therefore, lags may serve as a powerful tool for distinguishing between different spectral components or emission phases. Regarding the energy range 30 - 100 MeV covered by LLE data, we can think of GRB spectral evolution in the framework of a simple phenomenological scenario, in which we can distinguish two main cases:

\begin{itemize}
    \item The LLE spectrum is the high-energy extension of the prompt component, which evolves hard-to-soft over time. In this context, hard photons will precede soft photons, i.e. we have $\tau_{14} > 0$, by our definition; 

    \item The LLE spectrum highlights the presence of an additional and harder component with respect to the GBM one, while the latter is fading (or has already faded). In this case, the LLE component may reveal a long-lasting, delayed emission. Here, hard photons will lag behind soft ones, resulting in $\tau_{14} < 0$. 
\end{itemize}

In order to explore the link between the time lags and the spectral properties in the LLE band, we  investigate the possible correlation between $\tau_{14}$ and the spectral shape of the spectrum in the LLE energy range. In fact, if we refer to the $\nu F_{\nu}$ representation of a simple powerlaw spectrum, a flat spectral profile corresponds to a photon index  $\alpha \sim -2$ according to the notation of Eq.4. For the purposes of our discussion, we define "hard" a spectrum with $\alpha > -2$, whereas a spectrum is considered "soft" if $\alpha < -2$. Under this framework, and based on the scenario described above, considering that the typical high energy part of the prompt emission spectrum has photon index $<-2$, we should expect that GRBs in our sample with $\tau_{14} > 0$ exhibit a soft LLE spectrum, characterized by $\alpha_{\text{LLE}} < -2$ (i.e. LLE is consistent with the high-energy extension of the prompt phase). Conversely, GRBs for which $\tau_{14} < 0$ are expected to have hard LLE spectra, corresponding to values of $\alpha_{\text{LLE}} > -2$ (suggesting a possible additional component with respect to the extrapolation of the prompt emission MeV spectrum). However, the threshold of $\alpha_{\text{LLE}} \sim -2$ should be considered as indicative because it is not possible to define a specific value that allows us to separate exactly the two cases.

\begin{figure*}[!ht]
    \centering
    \includegraphics[width=0.9\textwidth]{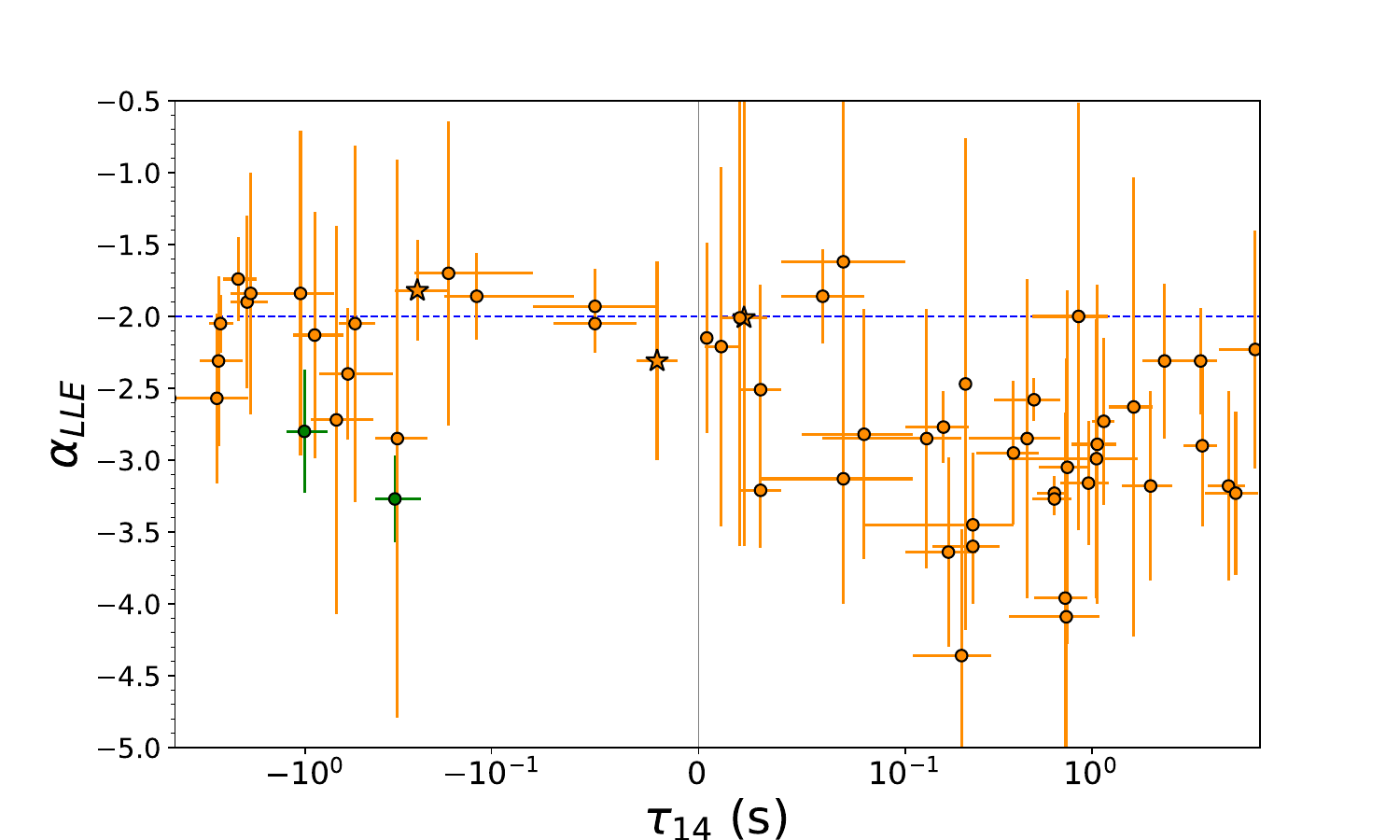} 
    \caption{LLE Photon Index as a function of the spectral lag between Band 1 (GBM-NaI) and Band 4 (LAT-LLE) for 56 out of 70 GRBs in our sample. Short GRBs are indicated by stars. Green points represent GRBs that deviate from the expected trend, i.e. with $\tau_{14} < 0$ and LLE Photon Index < -2 (considering the uncertainties on their values), corresponding to GRBs 100724B and 170405A. The dashed-blue line represent the arbitrary threshold between hard and soft spectra in the framework of the phenomenological scenario, described in Section \ref{Spectral Results}. Spectral lags $\tau$ are reported in symmetric-logarithmic scale for visualization purpose.}
    \label{lle_photon_index}
\end{figure*}

In Fig.~\ref{lle_photon_index} we compare the LLE spectral indices listed in Table \ref{table_spectral} with the corresponding values of $\tau_{14}$ from Table \ref{table_lags}. We observe that, for $\tau_{14} < 0$, the corresponding value of $\alpha_{\text{LLE}}$ tends to be approximately $\in$(-2,-2.5) or higher. Conversely, for $\tau_{14} > 0$, $\alpha_{\text{LLE}}$ becomes smaller than $\sim$ -2 in larger fraction of events despite its associated uncertainty makes it still consistent with -2.0. 

While for 54 out of the 56 analysed GRBs there is an agreement with the expectations of our simple scenario, there are two notable exceptions that deviate from this tendency, shown in green in Fig.~\ref{lle_photon_index}, corresponding to GRBs 100724B and 170405A. For these GRBs $\tau_{14}<0$ (at the $ \geq 2\sigma$ level), while $\alpha_{\text{LLE}}$ is significantly $< -2.0$  at $10\sigma$ and $6\sigma$, respectively.   

\renewcommand{\arraystretch}{1.2}
\begin{table*}[htbp]
\centering
\scalebox{0.84}{%
\begin{tabular}{l l l l l}
\hline
  GRB name & $\Delta t$ (s) & $\alpha_{\text{LLE}}$ & Norm ($10^{-8} \text{ergs/cm}^2\text{/s})$ & PG-Stat/DOF \\ \hline
  080825C & (-3.0, 37.0) & $-2.13^{+0.85}_{-0.86}$ & $3.55^{+0.12}_{-0.13}$ & 20.04/6\\
  080916C & (-3.2, 83.0) & $-2.05^{+0.20}_{-0.20}$ & $20.51^{+1.67}_{-1.17}$ & 2.45/6\\
  081224 & - & - & - & -\\
  090217A & (1.0, 41.0) & $-1.84^{+1.08}_{-1.17}$ & $3.04^{+1.33}_{-0.95}$ & 9.17/4\\
  090323A & (-15.0, 350.0) & $-3.45^{+0.46}_{-0.54}$ & $4.35^{+0.85}_{-0.77}$ & 6.69/4\\
  090328A & (5.0, 47.0) & $-3.18^{+0.61}_{-0.70}$ & $18.12^{+4.03}_{-4.38}$ & 2.88/4\\
  090902B & (-5.0, 54.0) & $-1.86^{+0.30}_{-0.30}$ & $25.26^{+ 2.18}_{-2.09}$ & 13.60/4\\
  090926A & (0.0, 72.0) & $-1.93^{+0.26}_{-0.26}$ & $33.86^{+2.61}_{-2.53}$ & 7.90/4\\
  091031A & (-3.0, 37.0) & $-2.23^{+0.89}_{-0.76}$ & $6.76^{+1.14}_{-1.37}$ & 6.19/4\\
  100116A & (70.5, 155.0) & $-2.85^{+1.34}_{-0.88}$ & $4.01^{+1.09}_{-1.42}$ & 8.63/4\\
  100225A & (2.6, 11.5) & $-2.72^{+1.49}_{-1.22}$ & $12.41^{+5.17}_{-5.08}$ & 3.47/6\\
  100724B & (-9.0, 92.0) & $-3.27^{+0.29}_{-0.31}$ & $19.86^{+1.87}_{-1.90}$ & 4.27/4\\
  100826A & (0.0, 137.0) & $-2.40^{+0.46}_{-0.45}$ & $14.57^{+1.19}_{-1.82}$ & 12.10/4\\
  110328B & (-8.0, 59.0) & $-2.99^{+1.14}_{-0.79}$ & $4.37^{+1.07}_{-1.20}$ & 6.73/4\\
  110721A & (-0.8, 24.0) & $-2.31^{+0.54}_{-0.52}$ & $19.73^{+2.92}_{-2.73}$ & 2.13/4\\
  110731A & (1.5, 12.8) & $-1.90^{+0.58}_{-0.62}$ & $7.46^{+1.18}_{-1.28}$ & 4.11/4\\
  120226A & (4.8, 13.7) & $-2.95^{+0.54}_{-0.48}$ & $6.54^{+2.85}_{-2.43}$ & 13.38/12\\
  120328B & (4.8, 13.7) & $-2.63^{+1.85}_{-1.41}$ & $133.06^{+52.93}_{-62.81}$ & 2.42/6\\
  120624B & (-6.0, 22.0) & $-2.85^{+2.47}_{-1.42}$ & $10.50^{+4.64}_{-10.48}$ & 20.38/4\\
  120709A & - & - & - & -\\
  121011A & - & - & - & -\\
  130305A & - & - & - & -\\
  130310A & (4.1, 4.4) & $-2.77^{+1.74}_{-1.68}$ & $527.37^{+278.85}_{-246.32}$ & 4.79/4\\
  130504C & (30.9, 32.1) & $-3.60^{+0.43}_{-0.36}$ & $2.93^{+1.05}_{-0.96}$ & 19.58/10\\
  130518A & (20.9, 23.7) & $-2.01^{+1.28}_{-1.70}$ & $2.04^{+0.88}_{-2.03}$ & 1.36/4\\
  130821A & - & - & - & -\\
  131014A & (-4.0, 20.0) & $-1.62^{+1.54}_{-1.21}$ & $16.09^{+8.42}_{-16.09}$ & 6.00/4\\
  131108A & (-1.0, 43.0) & $-1.86^{+0.33}_{-0.32}$ & $2.18^{+1.91}_{-1.82}$ & 5.36/4\\
  131216A & (-1.3, 4.3) & $-3.05^{+1.45}_{-1.00}$ & $3.78^{+3.14}_{-2.75}$ & 8.80/12\\
  131231A & - & - & - & -\\
  140102A & (1.5, 8.4) & $-2.41^{+0.61}_{-0.58}$ & $9.56^{+4.03}_{-3.57}$ & 15.86/10\\
  140206B & (-4.0, 74.0) & $-2.31^{+0.37}_{-0.36}$ & $14.99^{+1.52}_{-1.44}$ & 12.62/4\\
  140619B & - & - & - & -\\
  141028A & (-0.2, 58.0) & $-2.15^{+0.68}_{-0.64}$ & $7.17^{+1.25}_{-1.18}$ & 3.92/4\\
  141207A & (-3.0, 32.0) & $-1.84^{+0.87}_{-0.80}$ & $10.58^{+2.89}_{-3.17}$ & 8.71/4\\
  141222A & (-1.3, 2.4) & $-3.30^{+4.50}_{-2.52}$ & $3.38^{+3.64}_{-3.45}$ & 7.60/4\\
  150118B & - & - & - & -\\
  150202B & - & - & - & -\\
  150210A & (-5.0, 31.0) & $-2.07^{+1.33}_{-1.16}$ & $4.84^{+1.79}_{-2.61}$ & 10.00/4\\
  150403A & (-2.0, 28.0) & $-4.08^{+2.55}_{-2.05}$ & $1.23^{+1.70}_{-1.22}$ & 12.39/10\\
  150510A & (0.2, 18.0) & $-2.82^{+0.93}_{-0.82}$ & $7.15^{+3.11}_{-2.89}$ & 2.03/8\\
  150514A & - & - & - & -\\
  150523A & (-3.0, 45.0) & $-2.58^{+0.62}_{-0.56}$ & $1.65^{+0.65}_{-0.62}$ & 8.71/10\\
  150902A & (-3.0, 54.0) & $-1.70^{+0.95}_{-1.16}$ & $3.41^{+1.21}_{-1.14}$ & 13.82/6\\
  151006A & (-1.0, 45.0) & $-2.90^{+0.60}_{-0.51}$ & $4.54^{+1.38}_{-1.38}$ & 2.39/6\\
  160509A & (-76, 40.6) & $-2.58^{+0.15}_{-0.15}$ & $56.48^{+4.41}_{-4.32}$ & 3.06/6\\
  160625B & (184.0, 220.0) & $-3.23^{+0.12}_{-0.11}$ & $63.35^{+4.35}_{-4.28}$ & 6.51/6\\
  160709A & (-9.0, 38.0) & $-3.13^{+0.97}_{-0.77}$ & $3.11^{+1.46}_{-1.41}$ & 3.46/6\\
  160816A & (-0.04, 15.1) & $-2.21^{+1.31}_{-1.18}$ & $4.05^{+2.06}_{-2.06}$ & 15.31/6\\
  160821A & (1.2, 38.1) & $-3.27^{+0.11}_{-0.11}$ & $26.82^{+1.72}_{-1.71}$ & 2.84/6\\
  160905A & (-3.0, 46.0) & $-3.96^{+1.69}_{-0.89}$ & $1.54^{+1.16}_{-1.11}$ & 3.74/6\\
  160910A & (-3.0, 32.0) & $-3.64^{+0.70}_{-0.61}$ & $13.76^{+4.96}_{-4.61}$ & 11.47/6\\
  160917A & (0.9, 8.2) & $-2.85^{+0.97}_{-0.82}$ & $6.10^{+2.64}_{-2.82}$ & 2.61/8\\
  160917B & - & - & - & -\\
  170115B & (0.0, 35.0) & $-2.51^{+0.77}_{-0.69}$ & $3.80^{+1.34}_{-1.32}$ & 14.24/6\\
  170214A & (6.5, 86.5) & $-2.77^{+0.25}_{-0.24}$ & $11.40^{+1.49}_{-1.46}$ & 8.68/6\\
  170405A & (-7.0, 116.0) & $-2.80^{+0.45}_{-0.42}$ & $4.85^{+1.05}_{-1.04}$ & 6.10/6\\
  \hline
\end{tabular}%
}
\caption{Best fit parameters for the 66 long GRBs (upper part) and 4 short GRBs (lower part) in our sample. GRB name, time interval adopted for the spectral analysis (with respect to the LAT trigger time), spectral index ($\alpha_{\text{LLE}}$), normalization, and corresponding PG-Stat and DOF values. Errors are given at 90\% level. The spectral analysis is conducted in the 30 -- 100 MeV range. For GRBs 090227B, 120226A, 130504C, 131216A, 140102A, 150403A and 150523A, the lower energy bound is extended to 10 MeV.}
\label{table_spectral}
\end{table*}

\addtocounter{table}{-1}
\renewcommand{\arraystretch}{1.2}
\begin{table*}[htbp]
\centering
\scalebox{0.84}{%
\begin{tabular}{l l l l l}
\hline
  GRB name & $\Delta t$ (s) & $\alpha_{\text{LLE}}$ & Norm ($10^{-8} \text{ergs/cm}^2\text{/s})$ & PG-Stat/DOF \\ \hline
  170424A & - & - & - & -\\
  180720B & (-9.0, 45.0) & $-3.21^{+0.42}_{-0.38}$ & $13.28^{+3.07}_{-2.95}$ & 3.87/6\\
  190114C & (1.2, 38.1) & $-1.74^{+0.29}_{-0.29}$ & $119.47^{+136.66}_{-130.28}$ & 5.13/6\\
  190122A & (-1.5, 31.8) & $-3.16^{+0.46}_{-0.39}$ & $10.03^{+2.60}_{-2.58}$ & 2.64/6\\
  190531B & - & - & - & -\\
  190727B & (16.0, 29.0) & $-1.97^{+1.81}_{-1.36}$ & $4.73^{+2.82}_{-2.74}$ & 2.41/6\\
  190731A & (-13.0, 29.0) & $-2.89^{+1.25}_{-0.97}$ & $2.93^{+1.55}_{-1.93}$ & 6.01/6\\
  220101A & (80.5, 115.0) & $-3.23^{+0.61}_{-0.53}$ & $6.59^{+2.19}_{-2.06}$ & 5.97/6\\
  220209A & (4.0, 32.0) & $-2.73^{+0.58}_{-0.57}$ & $19.45^{+4.90}_{-4.57}$ & 3.75/6\\\\

  090227B & (0.2, 4.0) & $-2.31^{+0.69}_{-0.68}$ & $60.62^{+1.93}_{-2.14}$ & 8.97/10\\
  090510A & (-0.4, 23.0) & $-1.82^{+0.35}_{-0.34}$ & $26.27^{+2.56}_{-2.43}$ & 11.11/4\\
  110529A & - & - & - & -\\
  190606A & (-0.6, 0.9) & $-2.01^{+1.28}_{-1.70}$ & $2.04^{+1.03}_{-0.89}$ & 2.62/6\\
\hline
\end{tabular}%
}
\caption{Continued}
\end{table*}

To further test the reliability of the phenomenological scenario we defined above, it is useful to search for differences in the spectral behavior of the emission of the GRBs in our sample between the GBM range (10 - 1000 keV, i.e. where most of prompt emission occurs) and that in the LLE range (30 - 100 MeV), which we previously analysed for 56 GRBs in our sample (see Table \ref{table_spectral}).

\begin{figure*}[!ht]
    \centering
    \includegraphics[width=0.9\textwidth]{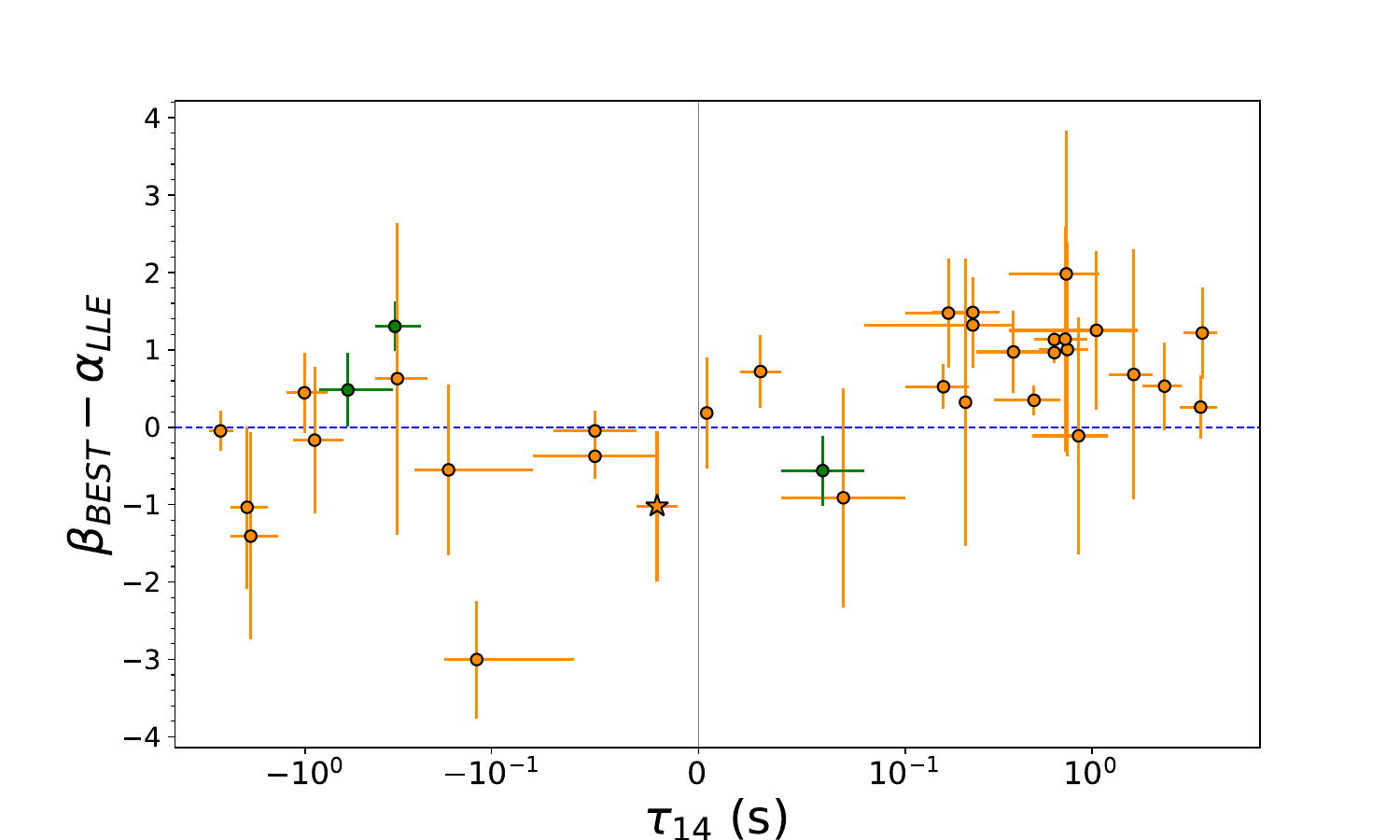} 
    \caption{Difference between high-energy photon index of the GBM emission ($\beta_{\text{BEST}}$) and the LLE Photon Index ($\alpha_{\text{LLE}}$) as a function of the spectral lag between Band 1 (GBM-NaI) and Band 4 (LAT-LLE) for 35 out of 70 GRBs in our sample. Short GRBs are indicated by stars. Green points represent GRBs that deviate from the expected trend, corresponding to GRBs 100724B, 100826A and 131108A. The dashed-blue line represent the arbitrary threshold between hard and soft spectra in the framework of the phenomenological scenario described in Section \ref{Spectral Results}. Spectral lags $\tau$ are reported in symmetric-logarithmic scale for visualization purpose.}
    \label{alpha_beta}
\end{figure*}

In particular, to investigate the difference between the GBM and LLE spectra and how does this comparison relate to the corresponding $\tau_{14}$ values, we compute the difference between their respective spectral indices $\beta_{\text{BEST}}$ and $\alpha_{\text{LLE}}$ as a function of $\tau_{14}$, as shown in Fig.\ref{alpha_beta}. In this context, $\beta_{\text{BEST}}$ represents the photon index that best fits the higher-energy segment of the GBM-BGO spectrum. The fitting model employed is either a Band function or a smoothly broken power law, depending on which provides the optimal fit for the specific GRB involved. We obtain values of $\beta_{\text{BEST}}$ for 35 out of the 70 GRBs in our sample.

This comparison shows a quite distinct behavior, in agreement with the one observable in Fig.\ref{lle_photon_index}. For negative time lags, the difference between the two spectral indices (both defined negative) tends to be greater than zero, indicating that $\alpha_{\text{LLE}} > \beta_{\text{BEST}}$, thus suggesting that the spectrum becomes harder at higher energies, such as those sampled by LLE data. However, within the negative-lag region, there is substantial scatter among the data points, and no conclusive statements can be made. For positive lag values, the difference between $\beta_{\text{BEST}}$ and $\alpha_{\text{LLE}}$ tends to fall below zero, meaning that  $\alpha_{\text{LLE}} < \beta_{\text{BEST}}$. In this positive lag region, the scatter among the data points diminishes, making this clustering more pronounced. 

We calculate the Spearman rank correlation coefficient between the spectral lags, $\tau_{14}$, and the difference $\beta_{\text{BEST}} - \alpha_{\text{LLE}}$ to search for linear correlation between the two datasets. The resulting correlation coefficient is 0.37 and a p-value of 3.2\% is obtained, corresponding to a $\sim 2.2\sigma$ confidence level. This indicates that the positive correlation between these two datasets, although quite weak, suggest a non-random and modest association between the two variables. Similarly to  Fig.~\ref{lle_photon_index}, some GRBs deviate from the expected trend of our proposed scenario, even considering their respective uncertainties. This is the case for GRBs 100724B (which is also an outsider in Fig.~\ref{lle_photon_index}), 100826A and 131108A, which are reported in green in Fig.~\ref{alpha_beta}. In this case, GRB 170405A, which was an outsider in Fig.~\ref{lle_photon_index}, behaves accordingly to what we expect from our simplified model.

\section{Discussion and Conclusions} \label{Conclusions}

In this work we presented the first systematic analysis of  \emph{Fermi}/LLE data to study spectral lags in GRBs. We analyze 66 long and 4 short GRBs with LLE data available in the online catalog, computing spectral lags across multiple energy bands. Our reference band is defined by the GBM-NaI detector (10–100 keV) and we compute how the emission at higher energy bands and in particular those sampled by the LLE data (30–100 MeV) lags with respect to this reference band (Fig.~\ref{schematic}). 

As emphasized by \cite{hakkila2008correlations}, being time-integrated properties, spectral lags and their uncertainties are mainly influenced by the light curve S/N and by the width of the DCF peaks. Nevertheless, lag computed over the full duration of a burst (which potentially comprises multiple emission episodes), are typically found to be dominated by the brightest and "spikiest" peak in the light curve (\citealt{mochkovitch2016simple}). In fact, this is the case for 96\% of the GRBs in our sample, with their DCF revealing a dominant peak (at all energies), resulting in the estimation of one spectral lag value. In the remaining cases, corresponding to GRBs 080916C, 090328A, 150523A, two values of $\tau_{14}$ were found (as shown in Fig.~\ref{all_lags}).

As shown in Table~\ref{table_lags}, long GRBs predominantly exhibit positive lags in lower-energy bands (98.6\% for $\tau_{12}$ and 94.3\% for $\tau_{13}$), consistent with previous studies interpreting positive lags as due to the hard-to-soft spectral evolution of the prompt emission (\citealt{daigne1998gamma, bocci2010lag, mochkovitch2016simple, bovsnjak2014spectral}). In contrast, $\tau_{14}$ values are more evenly split between positive (60\%) and negative (40\%) and are generally longer, averaging a few seconds. This suggests that at LLE energies, a delayed component distinct from the prompt emission may emerge when $\tau_{14} < 0$.

Short GRBs in our sample, apart from the peculiar GRB 090510A, exhibit spectral lags consistent with zero, in agreement with findings by \cite{yi2006spectral} and \cite{norris2006short}, the latter reporting that ~95\% of 260 short bursts in a BATSE sample show negligible lags. Similarly, \cite{guiriec2010time} observed minimal lags below 1 MeV in three bright short GRBs detected by Fermi, while \cite{bernardini2015comparing} found near-zero time delays in six short GRBs observed by Swift.

By comparing the three spectral lag values and their energy dependence, two main trends are identified. Around 40\% of GRBs display a decreasing lag with energy, transitioning from positive to negative at higher energies, indicative of a potential spectral inversion around 10–100 MeV. Notably, what we find for these GRBs is consistent with the results of \cite{bovsnjak2022grb}, who showed through simulations of single-pulse GRBs that the time lag with respect to the mean energy of the cross-correlation channel exhibits a specific trend: the light curves initially peak earlier as the energy increases, but this trend reverses above $\sim$10-100 MeV. Another 36\% show increasing lags with energy, while the remaining 24\% exhibit irregular patterns, likely due to low S/N or complex light curve structures.

We also investigate the relationship between spectral lags and key GRB prompt-emission features detected by GBM. We find that larger absolute lags are primarily associated with intermediate prompt-emission durations ($T90_{\text{GBM}} \sim$ 10–200 seconds) and intermediate values of the 10-1000 keV fluence. However, it is unclear whether these patterns arise from observational biases, selection effects, or intrinsic GRB properties. Further detailed studies are needed to disentangle these factors and better understand the underlying physics.

Lastly, we want to test if spectral lags evaluated across different energy bands can serve as a tool for distinguishing between different emission components. In order to do that, we start by building a simple phenomenological scenario in which the LLE emission (30 - 100 MeV) can be either identified as (i) the high-energy extension of the prompt component, evolving hard-to-soft over time (so that we should expect $\tau_{14} > 0$), or (ii) an harder, delayed component which arises as the prompt fades (implying $\tau_{14} < 0$). The harder spectral component in GRBs may arise from several mechanisms: Synchrotron Self-Compton processes in jets with strong magnetic fields (\citealt{panaitescu2000gamma, stern2004gamma}); high-energy photon production near the photosphere involving thermal, non-thermal, and magnetic reconnections (\citealt{ito2019photospheric, song2022origin}); forward shocks due to the interaction of the expanding jet with the interstellar medium, linked to high-energy afterglows (\citealt{kumar2010external, ghirlanda2010onset}); or hadronic models, such as proton-synchrotron emission and photohadronic cascades, which require substantial energy budgets (\citealt{pe2015physics, bovsnjak2022grb}).

To assess the reliability of such scenario, we compare $\tau_{14}$ values with the photon index $\alpha_{\text{LLE}}$ we obtain from the time-integrated spectral analysis of LLE spectra, fitting them with a simple power-law model (see Fig.~\ref{lle_photon_index}). Considering $\alpha_{\text{LLE}} \sim - 2$ as an indicative  threshold between an hard and a soft spectrum, GRBs with $\tau_{14} > 0$ are expected to have $\alpha_{\text{LLE}} < -2$, while $\tau_{14} < 0$ should correlate with cases in which $\alpha_{\text{LLE}} > -2$. Our results support this framework (with the two clear exception of GRBs 100724B and 170405A), with most GRBs hinting towards a correlation between $\tau_{14}$ and $\alpha_{\text{LLE}}$. Values of $\tau_{14} > 0$ may indeed be attributed to cases where the LLE emission is the extension of the prompt phase, evolving in a hard-to-soft fashion. This aligns with the observation that the vast majority of values for $\tau_{12}$ and $\tau_{13}$ are positive (98.6\% and 94.3\%, respectively). Transitions from positive to negative spectral lags can occur at LLE energies (around 10 MeV or higher) when a new, harder spectral component emerges. However, the large scatter in the $\tau_{14} < 0$ region, highlights possible challenges in interpreting negative spectral lags, due to limited LLE spectral statistics and the simplicity of the chosen fitting model. Estimating $\alpha_{\text{LLE}}$ is difficult due to poor data quality, and fitting with a simple power-law model may overlook complex spectral features, requiring additional components. The proposed phenomenological scenario is also a simplified representation that cannot fully capture more complex cases.

Notably, some GRBs in our sample for which $\tau_{14} < 0$ and $\alpha_{\text{LLE}} > -2$ exhibit an additional high-energy component. For example, the delayed > 100 MeV emission in GRB 080916C (with $\alpha_{\text{LLE}} \sim -2.05$) is attributed to proton synchrotron radiation during the prompt phase (\citealt{2010OAJ.....3..150R}). In contrast, for GRB 090510A, high-energy emission is linked to electron synchrotron radiation in the early afterglow phase (\citealt{kumar2010external, ghirlanda2010onset}). Similarly, delayed GeV emission in GRB 110731A likely corresponds to the high-energy tail of the afterglow (\citealt{ackermann2013multiwavelength}), and \cite{ravasio2019grb} interpreted the LAT emission in GRB 190114C as originating from the same afterglow-related mechanism. However, for GRBs like 090902B (\citealt{abdo2009fermi}) and 141207A (\citealt{arimoto2016high}), the origin of the high-energy component remains unclear, posing challenges for current models.

Taking $\beta_{\text{BEST}}$ as the representative photon index of the higher-energy segment of the GBM spectrum, we also focus on the difference between $\beta_{\text{BEST}}$ and $\alpha_{\text{LLE}}$, as a function of $\tau_{14}$ (see Fig.~\ref{alpha_beta}), in order to further test our framework. 
Values of $\beta_{\text{BEST}}$ are given for 35 GRBs considering either a Band function or a smoothly broken power law, depending on which was the best-fitting model as reported in the GBM Catalog.

The analysis reveals a behavior consistent with that seen in Fig.~\ref{lle_photon_index}. Overall, for positive values of $\tau_{14}$, the observed cluster aligns well with the hard-to-soft evolution typical of the prompt. As $\tau_{14}$ transits to negative values, a greater scatter among the data points occurs, making the interpretation less satisfactory and the clustering of the data more modest. We also see deviations from the expected trend for GRBs 100724B, 100826A, and 131108A. The linear correlation coefficient between $\tau_{14}$ and $\beta_{\text{BEST}} - \alpha_{\text{LLE}}$ is found to be 0.37 (with a p-value of 3.2\%, corresponding to a $\sim 2.2\sigma$ confidence level), suggesting a modest correlation between the two sets of data. 
   However, it is important to note that the large scatter observed, particularly in the region where $\tau_{14} < 0$, may be attributed to the limited quality of the LLE spectra and the lag estimates for the GRBs in our sample, or to more peculiar and complex features in their spectra, which we fit with a simple power-law model. 

This work emphasizes the value of spectral lag studies, especially across different energy ranges, as a tool for distinguishing between GRB emission components. Using LLE data to calculate time lags between 30–100 MeV provides insights into the onset of hard, time-delayed components and their interaction with the prompt emission. While spectral lag studies complement spectral analysis, further detailed models are needed, as LLE data alone lack sufficient statistics for deeper analysis.

In this work, we compute time lags by applying the DCF method to the entire interval of the LLE light curve in which emission is present, aiming to obtain a global measure of the time lag for each selected energy band. However, as noted by \cite{hakkila2004quiescent}, different emission episodes within a single GRB can exhibit distinct lags, suggesting that each episode may correspond to a different burst component. Moreover, previous studies have also indicated that GRBs can present both a slowly-varying ("smooth") component and a rapidly-varying ("spiky") component that overlap in the light curve (\citealt{cline1980detection, mitrofanov1989microsecond, norris1996attributes, walker2000gamma}). Our present analysis, based on time-integrated data, does not specifically address these individual components or their spectral evolution. Given the encouraging results presented in this work, a time-resolved study of lags in GRBs with multiple emission episodes will be the focus of a forthcoming paper.

These findings open avenues for deeper analysis, with spectral lag studies serving as a critical complement to broader multi-wavelength and multi-messenger efforts in understanding GRBs. Given the encouraging results presented in this work, in a forthcoming publication we aim to carry out a more in-depth study, consisting of both temporal and spectral analysis, for some specific GRBs from this sample, which represent great case-studies to assess all these interesting issues. 

\begin{acknowledgements}
We are thankful to M. G. Bernardini, F. Daigne and R. Mochkovitch for  valuable comments and discussion, which significantly contributed to shaping this work. C.M. is grateful to the Observatory of Brera - Merate for their warm hospitality during the course of this project. LN and GG acknowledge the funding support from the European Union-Next Generation EU, PRIN 2022 RFF M4C21.1 (202298J7KT - PEACE). AT acknowledges the "HERMES Pathfinder–Operazioni 2022-25-HH.0" grant.
\end{acknowledgements}

%Bibliography 
\bibliographystyle{aa} % style aa.bst
\bibliography{paper_new} % your references Yourfile.bib

\begin{appendix}
\label{appendix}
\onecolumn
\section{}

\begin{figure*}[ht]
    \centering
    \includegraphics[width=1\textwidth]{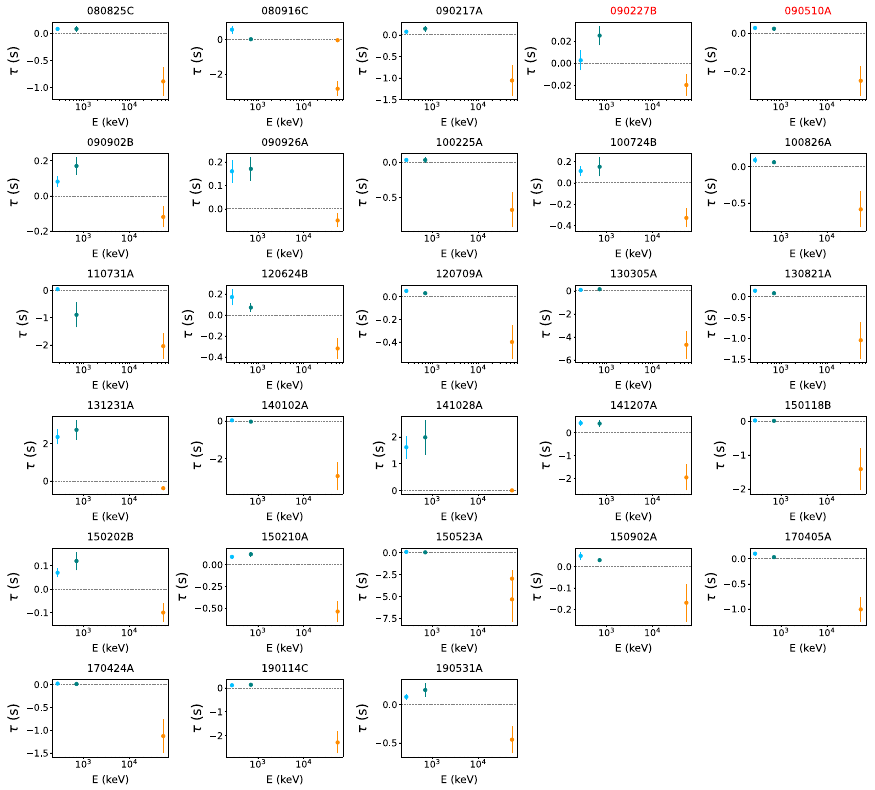} 
    \caption{28 GRBs in our sample (names in red indicate short GRBs) which exhibit a transition from positive values of $\tau_{12}$ and $\tau_{13}$ to a negative value of $\tau_{14}$. Band 1 - Band 2 (blue), Band 1 - Band 3 (green), and Band 1 - Band 4 (orange) spectral lags are compared with the mean energy $E$ of each channel, respectively.}
    \label{transition_grbs}
\end{figure*}

\begin{figure*}[!ht]
    \centering
    \includegraphics[width=1\textwidth]{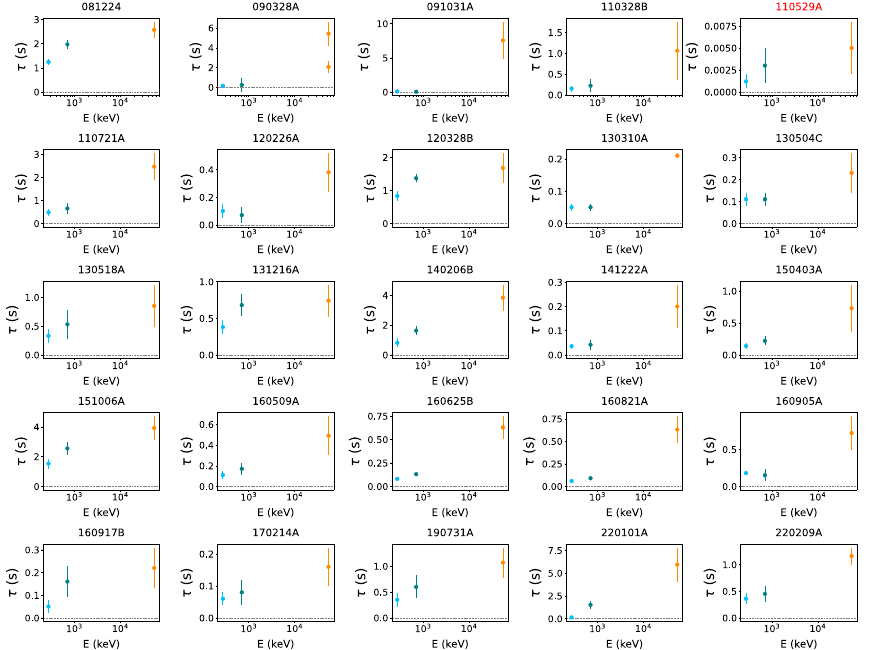} 
    \caption{25 GRBs in our sample (names in red indicate short GRBs), for which the spectral lags increase with higher values of $E$, remaining always positive. Band 1 - Band 2 (blue), Band 1 - Band 3 (green), and Band 1 - Band 4 (orange) spectral lags are compared with the mean energy $E$ of each channel, respectively.}
    \label{positive_grbs}
\end{figure*}

\begin{figure*}[!ht]
    \centering
    \includegraphics[width=1\textwidth]{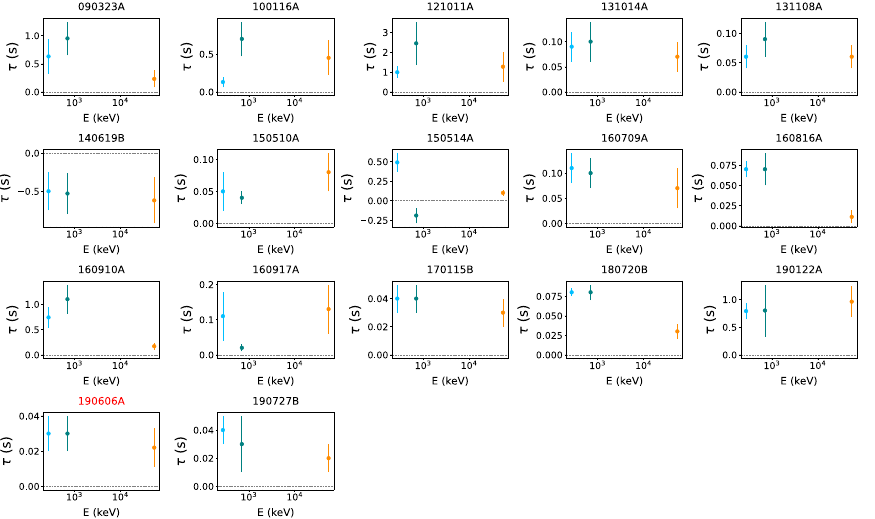} 
    \caption{17 GRBs in our sample (names in red indicate short GRBs) which exhibit more complex and various behaviors. Band 1 - Band 2 (blue), Band 1 - Band 3 (green), and Band 1 - Band 4 (orange) spectral lags are compared with the mean energy $E$ of each channel, respectively.}
    \label{hybrid_grbs}
\end{figure*}

\begin{figure*}[ht]
    \centering
    \includegraphics[width=1\textwidth]{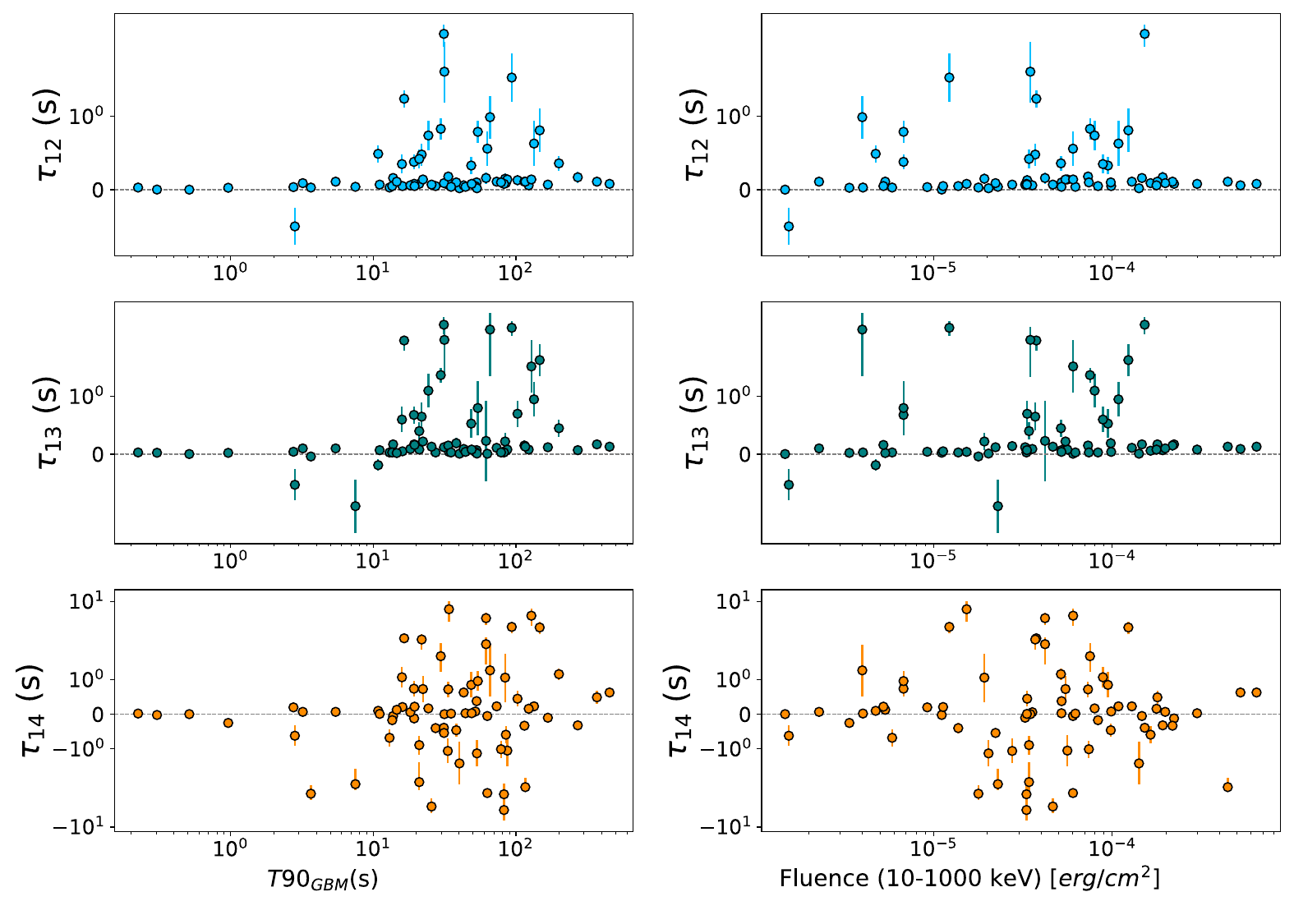} 
    \caption{Spectral lags as a function of the duration of the burst as seen by the GBM detector (left panel) and of the respective fluence (right panel). From top to bottom : Band 1 - Band 2 (blue points), Band 1 - Band 3 (green points), and Band 1 - Band 4 (orange points) spectral lags.}
    \label{t90_and_fluence}
\end{figure*}

\end{appendix}

\end{document}